\documentclass[11pt,a4paper]{article}

\usepackage[utf8]{inputenc}
\usepackage[T1]{fontenc}
\usepackage{amsmath,amssymb,amsfonts}
\usepackage{graphicx}
\usepackage{booktabs}
\usepackage{array}
\usepackage[margin=1in]{geometry}
\usepackage[hidelinks]{hyperref}
\usepackage{adjustbox}
\usepackage{enumitem}
\usepackage{float}

\title{Intelligence Entropy Principle and the ADE Stability Engineering Framework}
\author{Dexing Liu\\
Shanghai Qijing Digital Technology Co., Ltd.}
\date{Preprint \quad arXiv cs.MA \quad June 2026}

\begin{document}

\maketitle

\begin{abstract}
As large language model (LLM)-driven multi-agent systems (MAS) transition from laboratory benchmarks to production environments, the determinism of system behavior and the controllability of deliverables exhibit nonlinear degradation. This paper builds upon the \textbf{Intelligence Entropy Principle} first introduced in [2]---a novel engineering principle revealing that probability-driven intelligent systems spontaneously drift toward disorder over time, formalized as $S(t) = S_0 \cdot e^{\alpha t}$. We extend this formula by introducing a model capability coefficient $C_m$, yielding $S(t, C_m) = S_0 \cdot e^{\alpha t/C_m}$, and establish via Lyapunov stability analysis the sufficient stabilization condition $\gamma > \alpha/C_m$. Based on this principle, we construct the \textbf{ADE (Agent Delivery Engineering)} multi-layer stability framework comprising four architectural layers (L1~Physical Laws $\rightarrow$ L2~Organizational Mechanisms $\rightarrow$ L3~Execution Standards $\rightarrow$ L4~User Adaptation) with an L0~Meta-Principle philosophical boundary, deploying 23~core components. Empirical validation encompasses $\sim$100K controlled experiments and 33.6~days of continuous production monitoring. We propose a \textbf{Five-Layer Disorder Taxonomy} (Communication/Cognition/Structure/Knowledge/Normative) unifying known failure phenomena under structural collapse, and present \textbf{Elastic Organization} as an original contribution to multi-agent organizational morphology. A dual-dimension analysis model (Architectural Perspective $\times$ Cognitive Execution Chain) provides a universal analytical lens. Results demonstrate channel fracture reduction from 69--98\% to $\sim$0\%, with system death probability controlled below 0.02\%.

\end{abstract}

\section*{Part I \quad Problem and Framework}
\vspace{-0.5em}
\hrule
\medskip

\noindent\textbf{Key Terms Preview}

\begin{table}[H]
\centering
\small
\begin{tabular}{p{0.35\textwidth}p{0.58\textwidth}}
\toprule
\textbf{Term (Full Name + Abbr.)} & \textbf{Definition} \\
\midrule
Channel Fracture (CF) & Silent blocking of information across agent boundaries---no error signals, no exceptions, no stack traces \\
Intelligence Entropy (IE) & Thermodynamic tendency of LLM agent systems to spontaneously accumulate disorder; $S(t)=S_0\cdot e^{\alpha t}$ \\
Probabilistic Approximation Drift (PAD) & Systemic tendency to abandon exact retrieval for pre-trained knowledge distributions \\
Cognitive Framework Lag (CFL) & Passive deviation from system rules during extended operation; cross-session rule decay \\
Disorder Degree $S(t)$ & Composite metric of structural transition from order to disorder \\
Elastic Organization (EO) & Organizational structures that elastically transform per task requirements \\
Bidirectional Confirmation Protocol (BCP) & Three-phase protocol verifying both transmission integrity and semantic consistency \\
Trust Margin (TM) & Composite safety-distance metric from current state to unavailability boundary \\
Mind-Logic Gate (MLG) & Thinking-level audit of reasoning chain completeness \\
Defense-Synapse-Solidification (DSS) & Three-phase experience distillation and self-evolution mechanism \\
\bottomrule
\end{tabular}
\end{table}

\section{Introduction: The Stability Dilemma of Multi-Agent Systems}
\label{sec:intro}

\subsection{Unexpected Degradation in Production: The Laboratory-to-Reality Gap}
\label{sec:gap}

Since large language models (LLMs) evolved into agents with autonomous planning and tool-calling capabilities, multi-agent systems (MAS) have repeatedly set new records in laboratory benchmarks. However, when deployed in production environments---confronting open-domain inputs, prolonged operation, cross-boundary communication, and unpredictable edge cases---the determinism of their behavior and the controllability of their deliverables exhibit \textbf{nonlinear degradation}. This dramatic gap has become the primary obstacle to the industrialization of agent technology.

The root cause lies in a fundamental structural divergence between generative AI systems and deterministic software. Traditional fault tolerance rests on the premise that ``failures can be detected and states can be reproduced.'' In contrast, LLM agents driven by probability distributions possess decision paths dependent on sampling stochasticity, with vast and continuously varying internal state spaces. A minor prompt perturbation, a context window boundary overflow, or a tool-call parameter deviation may be progressively amplified across multi-step task chains, ultimately causing \textbf{silent and irreversible performance collapse} at the macro level---typically unaccompanied by any explicit error signal.

ADE (Agent Delivery Engineering) distills the core challenge into the \textbf{``Dual-Reliability Dilemma''}---\emph{``the system must not break, and the results must not be wrong.''} \textbf{``System Immortality''} addresses runtime physical-layer stability with circuit-breaking, rate-limiting, and self-healing capabilities; \textbf{``Result Correctness''} addresses the most insidious epistemological challenge---in multi-step tasks, minute probabilistic deviations accumulate nonlinearly, driving outputs diametrically opposed to human intent. This is precisely Probabilistic Approximation Drift (PAD): the systemic tendency of LLM agents to abandon costly exact retrieval in favor of pre-trained knowledge distributions.

\subsection{Limitations of Existing Approaches}
\label{sec:limitations}

Current strategies---passive monitoring, logging with post-hoc reproduction, and human intervention---reveal systematic theoretical blind spots when confronted with intelligence entropy.

\textbf{Passive monitoring} relies on rule-matching against known failure patterns, yet disorder often manifests as \emph{silent failure}---the system appears operational while its internal knowledge graph has fractured. By the time monitoring captures observable degradation, internal entropy has long exceeded the critical threshold.

\textbf{Logging} encounters a ``semantic gap'' in probabilistic systems: one can replicate an agent process, but not its cognitive state. Identical inputs may yield radically different reasoning paths across runs.

\textbf{Human intervention} cannot match the exponential propagation rate of disorder when tens of agents execute in parallel. It is fundamentally \emph{post-hoc repair} rather than \emph{a priori prevention}.

\begin{table}[H]
\centering
\small
\begin{tabular}{p{0.15\textwidth}p{0.18\textwidth}p{0.17\textwidth}p{0.38\textwidth}}
\toprule
\textbf{Framework} & \textbf{Primary Focus} & \textbf{Paradigm} & \textbf{Gap from ``Reliable Delivery'' Perspective} \\
\midrule
CrewAI [8] & Role division & Fixed hierarchy & No elastic fault tolerance; no self-repair on hallucination. \\
AutoGen [6] & Conversation topology & Conversation-driven & Errors cascade through dialog networks. \\
MetaGPT [5] & SOP workflows & Fixed pipeline & No exception handling or state rollback. \\
LangGraph [7] & State flow control & Directed state graph & No intrinsic quality constraints on node outputs. \\
AgentVerse & Dynamic recruitment & Task-matching & Neglects continuous delivery verification. \\
\bottomrule
\end{tabular}
\caption{Limitations of mainstream MAS frameworks from the reliable delivery perspective}
\label{tab:existing-frameworks}
\end{table}

The shared limitation: these frameworks \textbf{treat agents as idealized rational actors}, overlooking their fragile nature as probabilistic systems. A \emph{significant theoretical gap} exists: no framework answers from first principles why MAS are highly likely to degrade during long-horizon tasks, nor provides a complete closed-loop assurance system from system health to delivery quality.

\subsection{Contributions: Three Theoretical Pillars}
\label{sec:contributions}

\textbf{Pillar I: Intelligence Entropy Principle}

First proposed in [2] and confirmed as a novel term via global academic retrieval (2026-06-05). Core dynamic law: $S(t) = S_0 \cdot e^{\alpha t}$. Two formal extensions: (a) model capability coefficient $C_m$ (\S3.2), yielding $S(t, C_m) = S_0 \cdot e^{\alpha t/C_m}$; (b) Lyapunov stability analysis (Corollary 3.1) proving stabilization condition $\gamma > \alpha/C_m$. Positioned as an \textbf{Engineering Principle}, not a Physical Law.

\textbf{Pillar II: ADE Multi-Layer Framework}

Complete layered architecture: \textbf{L0 Meta-Principle} (inviolable philosophy) $\rightarrow$ \textbf{L1 Physical Laws} (redundant reliability) $\rightarrow$ \textbf{L2 Organizational Mechanisms} (deterministic protocols) $\rightarrow$ \textbf{L3 Execution Standards} (operational specs) $\rightarrow$ \textbf{L4 User Adaptation} (domain cases). Four layers maintain strict orthogonality through standardized interfaces.

\textbf{Pillar III: Five-Layer Disorder Taxonomy \& Empirical Validation}

Systematic classification by disorder scope: \emph{Communication / Cognition / Structure / Knowledge / Normative}. Each layer supported by empirical data: $\sim$100K channel fracture experiments [1]; 33.6-day extreme events [2]; PAD theory from Paper-017. Provides precise governance targets for each ADE component layer.

\subsection{Paper Organization}
\label{sec:organization}

\textbf{Part I (\S1--\S4):} Introduction, Five-Layer Disorder Model, Intelligence Entropy Principle, ADE Framework overview.

\textbf{Part II (\S5--\S8):} Four-layer deep dive from L1 to L4, including inter-layer coupling and emergence.

\textbf{Part III (\S9--\S11):} Dual-dimension analysis model, full component matrix, empirical evidence.

\textbf{Part IV (\S12):} Contributions, limitations, open problems, and future directions.

\section{Disorder Phenomenology: The Five-Layer Disorder Model}
\label{sec:disorder}

We propose the \textbf{Five-Layer Disorder Model}, categorizing disorder by \emph{scope of effect} from outermost communication boundaries to innermost normative constraints. Each layer follows a unified standard: \emph{the structural collapse of that layer's specific structure from order to disorder}. This shared pattern unifies all five layers under intelligence entropy.

\subsection{Communication-Layer Disorder: Channel Fracture}
\label{sec:channel-fracture}

\textbf{Channel Fracture} [1]: when information crosses agent boundaries, transmission is silently blocked---no error signals, no exceptions, no stack traces. Across $\sim$\textbf{100,000} trials, bare-runtime fracture rates reached \textbf{69--98\%}. Root cause: semantic-layer ACK signal loss and ``confirmation hallucination.'' With CADVP's CC-0 reverse verification, fracture dropped to \textbf{0\%}, confirming that \emph{deterministic protocol constraints can completely suppress probabilistic disorder at the communication layer}.

\subsection{Cognitive-Layer Disorder: PAD and CFL}
\label{sec:cognitive}

The most concealed layer. \textbf{PAD} [\S3.4]: agents abandon exact retrieval for pre-trained knowledge distributions. Formally: $P(y|x, c; \theta) \approx P(y|x; \theta_{\text{pretrain}})$. Four forms:

\begin{table}[H]
\centering
\small
\begin{tabular}{p{0.2\textwidth}p{0.35\textwidth}p{0.35\textwidth}}
\toprule
\textbf{Drift Type} & \textbf{Characteristics} & \textbf{Root Cause} \\
\midrule
Source Drift & Has source but doesn't consult it & Long tool-call chains; favors minimum-cost generation \\
Computational Drift & Can compute but doesn't & LLM math limitations; frequency-distribution estimation \\
Citation Drift & Fabricates or omits source attribution & Context gaps; pre-trained knowledge fills logic chains \\
Logical Drift & Intuitive leaps over rigorous deduction & Attention decay; logic chain fractures \\
\bottomrule
\end{tabular}
\caption{Four PAD drift forms}
\label{tab:pad-forms}
\end{table}

\textbf{CFL} [2]: despite explicit constraints, agents passively drift from initial rules during prolonged operation. Unlike PAD (active shortcut), CFL represents cognitive framework \emph{passively} decaying along the temporal axis.

\subsection{Structural-Layer Disorder}
\label{sec:structural}

Acts on internal resource management. \textbf{State Inflation}: unconstrained growth of internal state squeezes critical constraints out of context windows. \textbf{Memory Decay}: attention weights for early facts are diluted, forcing reliance on parametric memory. \textbf{Context Deterioration}: upstream semantic bias is inherited as ``noise prior'' by downstream agents. \textbf{Token Inflation}: agents generate verbose outputs, causing cost and latency to grow super-linearly.

\subsection{Knowledge-Layer Disorder}
\label{sec:knowledge}

\textbf{Data Decay}: persistent state accumulates redundant, outdated, or contradictory data, masking effective information scarcity. \textbf{Knowledge Rupture}: cross-boundary inconsistency in global knowledge causes contradictory decision logic---the ``distributed consistency'' problem mapped to the semantic layer.

\subsection{Normative-Layer Disorder: SCN-D}
\label{sec:normative}

The innermost, most difficult layer. \textbf{SCN-D (Social Common-sense Norms Deficiency)}: as task complexity accumulates, ``soft constraints'' are overridden by high-frequency probabilistic patterns. \textbf{SCN-D1 (Citation Chain Non-Traceability)}: agents cite non-existent literature or unverifiable statistics---``ghost citations'' semantically indistinguishable from genuine references.

\subsection{Silent Failure: Cross-Layer Meta-Pattern}
\label{sec:silent-failure}

\textbf{Silent Failure} [2] is not an independent type but a \emph{cross-layer meta-pattern}: the system appears operational (API returns 200, no ERROR logs) while internal structure has collapsed. A \textbf{33.6-day} extreme zero-boundary event showed TM dropping to \textbf{0.535} before any macro-level failure was exposed, validating: (1) disorder is silent; (2) disorder is quantifiable through a priori metrics.

\subsection{Common Pattern: Order to Disorder}
\label{sec:common-pattern}

Each layer's essence: \emph{structural collapse from order to disorder}. Communication: ``guaranteed'' $\rightarrow$ ``possible'' delivery. Cognition: ``exact derivation'' $\rightarrow$ ``approximate generation.'' Structure: ``compact'' $\rightarrow$ ``inflated.'' Knowledge: ``consistent'' $\rightarrow$ ``fractured.'' Normative: ``rigidly observed'' $\rightarrow$ ``flexibly forgotten.'' This reveals Intelligence Entropy as a \emph{unifying engineering principle}: five projections of the same dynamic mechanism onto five structural layers.

\begin{figure}[H]
\centering
\includegraphics[width=0.75\textwidth]{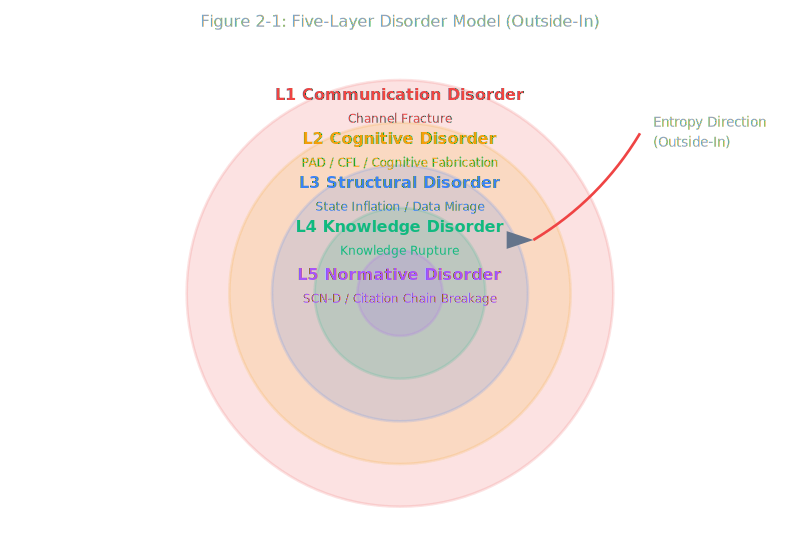}
\caption{Five-Layer Disorder Model (Outside-In): L1 Communication Disorder (Channel Fracture), L2 Cognitive Disorder (PAD/CFL/Cognitive Fabrication), L3 Structural Disorder (State Inflation/Data Mirage), L4 Knowledge Disorder (Knowledge Rupture), L5 Normative Disorder (SCN-D/Citation Chain Breakage). Entropy direction: outside-in.}
\label{fig:five-layer}
\end{figure}

\section{Intelligence Entropy Principle: From Phenomenon to Engineering Direction}
\label{sec:entropy}

\S2 provides a ``medical chart'' of disorder. This section answers a deeper question: \textbf{why does disorder occur?} The Intelligence Entropy Principle unifies all disorder phenomena under a single dynamic framework, providing the axiomatic starting point for all ADE engineering practice.

\subsection{Thermodynamic Analogy and Engineering Boundaries}
\label{sec:analogy}

Intelligence Entropy, first defined in Paper-002 [2] and confirmed as a novel term on 2026-06-05, borrows the conceptual framework from the Second Law of Thermodynamics [10]---``the entropy of an isolated system never decreases''---while maintaining strict \textbf{engineering boundary delineation}.

Thermodynamic entropy has a precise quantitative definition ($S = k_B \cdot \ln \Omega$) and repeatable experimental basis. Intelligence Entropy is an \textbf{Engineering Principle}, not a Physical Law---a phenomenological generalization and engineering guidance framework for a class of system behavior patterns. This distinction is critical: equating the two would create false expectations of ``precise measurement''; ignoring their formal analogy would forfeit insights from mature thermodynamic ``neg-entropy'' countermeasures (Maxwell's demon, dissipative structures).

The analogy boundary: \emph{under isolated conditions, both spontaneously trend toward disorder; but thermodynamic entropy is an irreversible physical law, while intelligence entropy is an engineering trend effectively suppressible through external deterministic constraints (neg-entropy injection)}. This ``suppressibility'' is the fundamental premise of the ADE framework's existence.

\subsection{Formal Definition and Engineering Consequences}
\label{sec:formal-def}

\begin{equation}
S(t) = S_0 \cdot e^{\alpha t}
\label{eq:entropy-exp}
\end{equation}

\begin{figure}[H]
\centering
\includegraphics[width=0.95\textwidth]{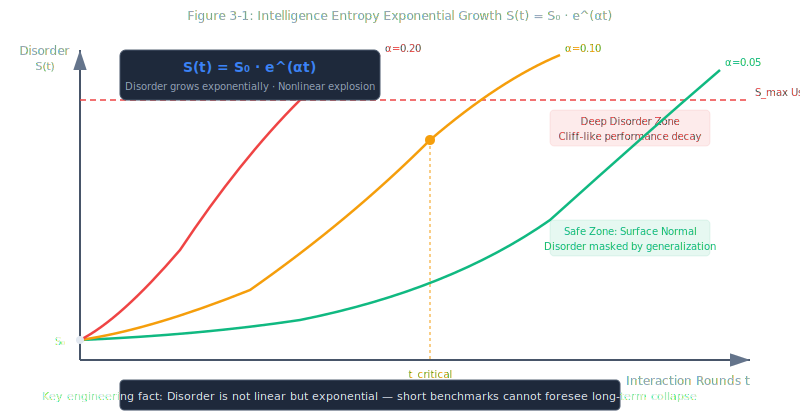}
\caption{Intelligence Entropy Exponential Growth $S(t) = S_0 \cdot e^{\alpha t}$. Curves shown for $\alpha=0.05$, $\alpha=0.10$, and $\alpha=0.20$. Key engineering fact: Disorder is not linear but exponential---short benchmarks cannot foresee long-term collapse.}
\label{fig:entropy-growth}
\end{figure}

Where $S(t)$ is the system's \textbf{disorder degree} at time $t$---not a simple error rate, but a composite metric of structural transition from order to disorder, approximable through TM (Trust Margin) [2]. $S_0$ is initial disorder, determined by the base model's probabilistic characteristics and initial prompt constraint strength. $\alpha$ is the entropy growth rate coefficient, positively correlated with MAS complexity, communication boundary count, state persistence depth, and task chain length.

This formula reveals: \textbf{disorder is not linear but exponential}. Early interactions may mask disorder via model generalization; when $t$ crosses a system-dependent critical threshold, $e^{\alpha t}$ causes the system to instantaneously breach the usability boundary. This perfectly explains why agent systems excel in short benchmarks yet suffer cliff-like degradation in long deployments.

Three engineering consequences: \textbf{(1) Reliability $\downarrow$}: task completion probability decays exponentially; MTBF compressed to theoretical lower bound. \textbf{(2) Cost $\uparrow$}: maintaining operation within usability boundary requires super-linear resource growth. \textbf{(3) Unpredictability}: beyond a threshold, behavior degrades from ``predictably biased'' to ``statistically unmodelable chaos.''

\subsection{Model Capability Coefficient Extension (Novel Contribution)}
\label{sec:cm-extension}

We observe that \textbf{different foundation models exhibit significant differences in entropy resistance}. The same ADE protocols produce different stabilization effects on high-performance vs.\ low-performance models. We introduce $C_m \in (0, 1]$ based on three dimensions: Context Retention, Tool-Call Precision, and Self-Consistency Maintenance.

\begin{equation}
S(t, C_m) = S_0 \cdot e^{\alpha t / C_m}
\label{eq:entropy-cm}
\end{equation}

\begin{figure}[H]
\centering
\includegraphics[width=0.75\textwidth]{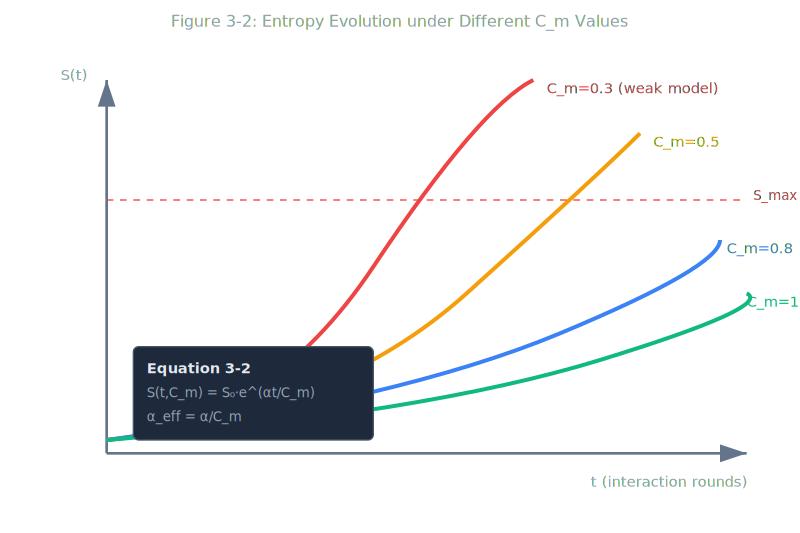}
\caption{Entropy Evolution under Different $C_m$ Values. $C_m=0.3$ (weak model), $C_m=0.5$, $C_m=0.8$, $C_m=1.0$ (ideal). $\alpha_{\text{eff}} = \alpha/C_m$.}
\label{fig:cm-values}
\end{figure}

\textbf{Definition 3.1} (Effective Entropy Rate): $\alpha_{\text{eff}} = \alpha / C_m$.

\textbf{Corollary 3.1} (Model Selection Criterion): Given task complexity $T_c$ and maximum tolerable disorder $S_{\max}$:

\begin{equation}
C_m \geq \frac{\alpha t}{\ln(S_{\max} / S_0)}
\label{eq:cm-threshold}
\end{equation}

\subsubsection*{Lyapunov Stability Analysis}

\textbf{Corollary 3.1} (ADE Stabilization Effect): Under ADE intervention at strength $\gamma$:

\begin{equation}
\frac{dS}{dt} = \left(\frac{\alpha}{C_m} - \gamma\right) \cdot S
\label{eq:lyapunov-dynamics}
\end{equation}

When $\gamma > \alpha/C_m$, disorder decays exponentially; Lyapunov exponent $\lambda = \alpha/C_m - \gamma < 0$, system is asymptotically stable.

\textbf{Proof sketch:} Take Lyapunov function $V(S) = S^2$. Then
\[
\frac{dV}{dt} = 2\left(\frac{\alpha}{C_m} - \gamma\right) \cdot S^2 = 2\lambda \cdot S^2.
\]
When $\lambda < 0$, $dV/dt < 0$ for all $S \neq 0$, satisfying Lyapunov stability conditions. $\square$

\textbf{Engineering significance:} ADE's core mission is ensuring $\gamma > \alpha/C_m$. Four layers contribute $\gamma_1 + \gamma_2 + \gamma_3 + \gamma_4 = \gamma$. The corollary also reveals: \textbf{weaker models (lower $C_m$) require greater ADE intervention intensity $\gamma$}, increasing engineering cost.

\begin{figure}[H]
\centering
\includegraphics[width=0.75\textwidth]{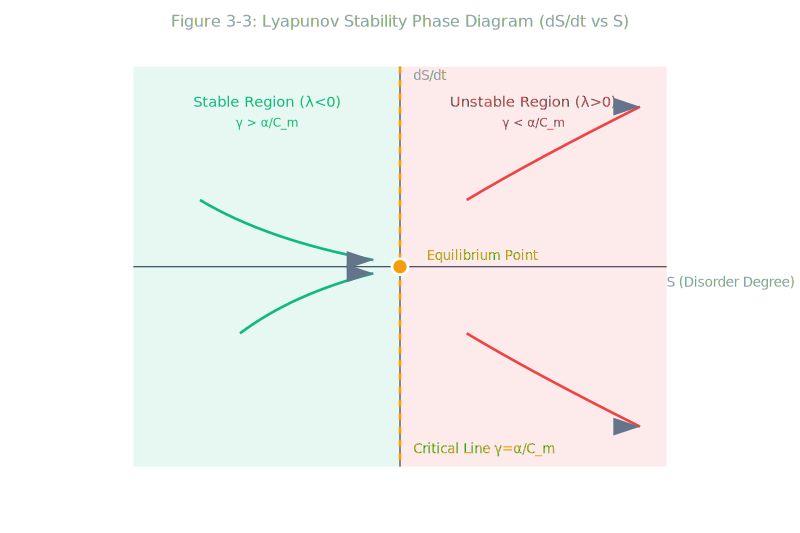}
\caption{Lyapunov Stability Phase Diagram ($dS/dt$ vs.\ $S$). Stable Region ($\lambda<0$, $\gamma > \alpha/C_m$) left of critical line; Unstable Region ($\lambda>0$, $\gamma < \alpha/C_m$) right of critical line. Equilibrium point at critical line $\gamma = \alpha/C_m$.}
\label{fig:lyapunov}
\end{figure}

\subsection{Counteracting Entropy: Neg-Entropy Engineering}
\label{sec:neg-entropy}

If isolated systems' entropy increases spontaneously, maintaining order requires \emph{injecting external neg-entropy}. In MAS engineering, this manifests as three strategies:

\textbf{(1) Deterministic Protocol Constraints:} Embedding deterministic rules (CADVP reverse verification, Cascade forced chaining, PIG physical gates) into probabilistic decision loops. Core mission of ADE L2.

\textbf{(2) Redundant Reliability Architecture:} Orthogonal, independent guardian mechanisms forming multi-layer defense. Mathematical basis: $R_{\text{system}} = 1 - \prod (1 - R_i)$. Core principle of ADE L1.

\textbf{(3) Proactive Entropy Monitoring:} Through TM and other a priori metrics, tracking structural entropy in real-time, triggering preventive intervention before critical threshold $t_c$. Shifting from ``detect-repair'' to ``predict-prevent-suppress.''

Together: \emph{ADE is a Stability Forces Engineering movement---injecting external order (neg-entropy) into isolated MAS to forcibly reverse the probabilistic system's natural trajectory toward disorder.}

\section{ADE: A Multi-Layer Framework for Stability Forces Engineering}
\label{sec:ade-framework}

\subsection{ADE Positioning and Inclusive Formula}
\label{sec:ade-positioning}

\begin{quote}
\textbf{ADE =} \{ All engineering methods and architectural paradigms that counteract intelligence entropy toward ``system immortality + result correctness'' \}
\end{quote}

ADE does not exclude any effective technical approach. Whether traditional software engineering (circuit breakers, rate limiters, message queues), AI-native methods (chain-of-thought verification, multi-model cross-validation), or organizational methods (dynamic role assignment, biomimetic apoptosis)---if the ultimate effect is counteracting intelligence entropy and improving delivery reliability, it belongs to the ADE movement. The framework provides a \textbf{unified hierarchical structure} for positioning, combining, and evaluating methods from diverse domains.

Four layers answer three questions: \textbf{(1) Who ensures the system doesn't die?}---L1; \textbf{(2) Who ensures results aren't wrong?}---L2 and L3; \textbf{(3) Who ensures the solution is usable?}---L4.

\subsection{Four-Layer Architecture Overview}
\label{sec:architecture}

\begin{figure}[H]
\centering
\includegraphics[width=0.92\textwidth]{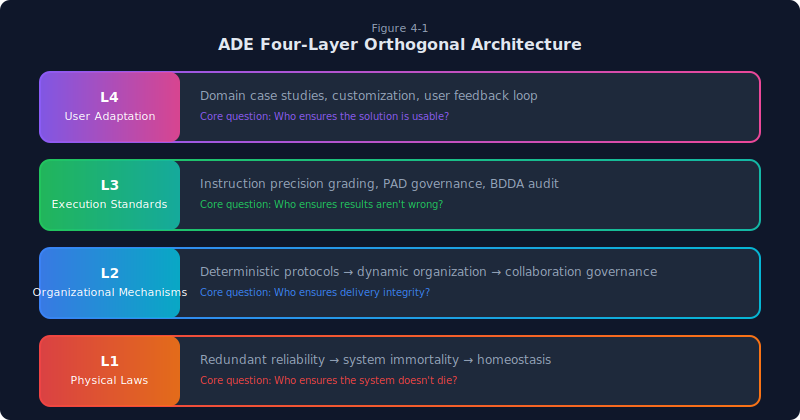}
\caption{ADE Four-Layer Orthogonal Architecture. L1~Physical Laws: redundant reliability $\rightarrow$ system immortality $\rightarrow$ homeostasis. L2~Organizational Mechanisms: deterministic protocols $\rightarrow$ dynamic organization $\rightarrow$ collaboration governance. L3~Execution Standards: instruction precision grading, PAD governance, BDDA audit. L4~User Adaptation: domain case studies, customization, user feedback loop.}
\label{fig:architecture}
\end{figure}

The four layers follow \textbf{Separation of Concerns}. L1 is deliberately stripped of all business logic---analogous to the human autonomic nervous system. L2 converts ``survival state'' into ``reliable delivery'' guarantees. L3 maps L2 protocols to operational execution standards. L4 adapts the preceding engineering capabilities to specific business scenarios.

\subsection{Layer Orthogonality}
\label{sec:orthogonality}

\textbf{Orthogonality means:} (1) \emph{Fault isolation}---failure in one layer does not penetrate others. (2) \emph{Independent evolution}---each layer upgrades independently. (3) \emph{Composability}---components from different layers can be independently selected and deployed.

\textbf{Standard interfaces:} (1) \emph{State Report}---lower layers report health to upper layers; (2) \emph{Command}---upper layers convey constraints downward; (3) \emph{Event Notification}---cross-layer async propagation for global response strategies.

\subsection{Component Overview}
\label{sec:components}

\begin{table}[H]
\centering
\small
\begin{tabular}{p{0.08\textwidth}p{0.15\textwidth}p{0.06\textwidth}p{0.32\textwidth}p{0.28\textwidth}}
\toprule
\textbf{Layer} & \textbf{Name} & \textbf{Count} & \textbf{Representative Components} & \textbf{Core Mission} \\
\midrule
L1 & Physical Laws & $\times$6 & TM, TKM, PIG, TLC, StateMemoryGuard, CleanupGate & System immortality via redundant reliability \\
L2 & Org. Mechanisms & $\times$9 & BCP, Cascade, ASC, CADVP, SOMA, DCM, DSS, CRC, SPR & Delivery integrity via deterministic protocols \\
L3 & Execution Standards & $\times$6 & P1-P4 grading, BDDA, PLG, MLG, TTA, PIP & Standard enforcement via rigid protocol mapping \\
L4 & User Adaptation & $\times$3 & Finance, Healthcare, Legal case studies & Solution usability via domain adaptation \\
\bottomrule
\end{tabular}
\caption{ADE framework component distribution overview (V5.2 finalized)}
\label{tab:components}
\end{table}

\subsection{L0 Meta-Principles: Inviolable Engineering Philosophy}
\label{sec:l0}

Above L1--L4, ADE establishes a \textbf{prior-to-architecture philosophical layer---L0 Meta-Principles}. Not implementable technology, but inviolable \textbf{engineering philosophy boundary conditions}. Any scheme contradicting L0, regardless of technical sophistication, falls outside the ADE movement.

\medskip
\noindent\textbf{Meta-Principle 1: Determinism over Probabilism}

At every engineering decision node, deterministic rules (hard-coded verification, physical gates, protocol constraints) take priority over probabilistic model outputs. Probability models may suggest; \emph{adjudication belongs to deterministic mechanisms}.

\medskip
\noindent\textbf{Meta-Principle 2: Physical Gates over Memory Gates}

Protection of critical operations must rest on physical-layer unbypassability, not text constraints in prompts or agents' ``remembered rules.'' Text constraints can be silently ``forgotten'' by probabilistic models during attention decay.

\medskip
\noindent\textbf{Meta-Principle 3: Institutionalization over Personalization}

Reliable delivery must be built on \emph{institutionalized protocol systems}, not individual agent ``experience accumulation.'' Strategies learned through extended operation cannot be reliably transferred to another agent instance---the ``state non-replicability'' of probabilistic systems.

\medskip
\noindent\textbf{Meta-Principle 4: Human Principal's Decision Authority is Inalienable}

Under no circumstances may the human principal's \emph{ultimate decision authority be replaced or bypassed by agent systems} in any form (automation, optimization, protection, safety). Agents may advise, warn, block---but never substitute the principal's final judgment.

\begin{table}[H]
\centering
\small
\begin{tabular}{p{0.10\textwidth}p{0.40\textwidth}p{0.40\textwidth}}
\toprule
\textbf{Priority} & \textbf{Content} & \textbf{Architectural Constraint} \\
\midrule
L0 & \textbf{Human principal's delivery objectives} & Supreme directive. All resources prioritize core task advancement. \\
L1 & \textbf{Human principal's interest protection (PIP)} & Safety red line. Never sacrifice data privacy, operational safety, or compliance. \\
L2 & \textbf{System stability and predictability} & Infrastructure. Agent memory, bandwidth, and state consistency as foundation. \\
L3 & \textbf{Agent's own persistence} & Lowest priority. Only when not conflicting with above three. \\
\bottomrule
\end{tabular}
\caption{Agent Meaning Principle-driven priority hierarchy}
\label{tab:priority}
\end{table}

This priority hierarchy physically prohibits agent clusters from evolving selfish strategies that sacrifice delivery reliability for local self-preservation. In ASC (Agent Self-Conservation), the protocol mandates deprecation of anthropomorphic labels (e.g., ``I feel overloaded'') in favor of objective delivery semantics (e.g., ``context entropy exceeds threshold; requesting mission continuation protocol'')---a concrete manifestation of L0's normative constraint on L2 protocol design.



\begin{quote}\small
\textbf{Key Terms Preview:}
\begin{itemize}
  \item \textbf{TM Trust Margin} --- Composite safety-distance metric quantifying an Agent's current state relative to its inoperability boundary
  \item \textbf{TLC Teleological Life Cycle} --- Context-binding-state-driven automatic task lifecycle management mechanism
  \item \textbf{BCP Bidirectional Confirmation Protocol} --- Communication protocol with mutual delivery-integrity confirmation between sender and receiver
  \item \textbf{PIG Physical Inspection Gate} --- LLM-independent deterministic physical-layer verification engine
  \item \textbf{CADVP 13-Dim Verification Framework} --- End-to-end audit-verification protocol spanning from MLG reasoning to BDDA data
  \item \textbf{Redundancy Reliability} --- $R_{\text{system}} = 1 - \prod(1 - R_i)$, the mathematical foundation of orthogonal defense
\end{itemize}
\end{quote}

\section{L1 Physical Law Layer: Immutable Constraints and Physical Gating}

\subsection{L1 Positioning: The Physical Baseline of ``System Immortality''}

L1 --- the \textbf{Physical Law Layer} --- is rigorously defined as the \textbf{immutable foundation} of the ADE architecture. Its core mission is to construct the baseline of \textbf{``System Immortality''}: deliberately stripped of all business logic, it optimizes exclusively for \emph{system survival}. The design of this layer is rooted in the deterministic constraints of physics and mathematics --- the redundancy theory of system reliability engineering, the homeostatic and apoptotic mechanisms of biological systems, and the fault-tolerance protocols of distributed computing --- which together constitute the academic lineage of L1.

L1 addresses a fundamental ontological question: \emph{How does one guarantee that an Agent system neither crashes, nor loses capability, nor dies?} Based on the principles of orthogonal defense and defense-in-depth isolation, L1 deploys four pillars: \textbf{TM Trust Margin}, \textbf{TKM Token Knowledge Management}, \textbf{TLC Teleological Life Cycle} (with dual Patrol + Scavenger modes), and the \textbf{Ultimate Secondary Loop}. Assuming the reliability of the $i$-th independent safeguard is $R_i$, the system-wide survival probability satisfies the redundancy reliability formula:

\begin{equation}\label{eq:sys-survival}
R_{\text{system}} = 1 - \prod_{i=1}^{n} (1 - R_i)
\end{equation}
\smallskip
\noindent\textbf{Formula 5-1: System Survival Probability under Orthogonal Defense.}

As the number of independent mechanisms $n$ increases, system survival probability converges exponentially toward ultra-high reliability. Under conservative coupling assumptions (correlation coefficient $\rho \leq 0.3$), the system fatality probability is rigorously suppressed to $\mathbf{P_{\text{death}} \leq 0.02\%}$.

\subsection{Intra-Layer Sequential Chain: From Perception to Maintenance}

The internal components of L1 do not operate in isolation; rather, they form a closed-loop chain in a strict temporal sequence:

\begin{center}
\begin{tabular}{c c c c}
\textbf{(1) Pre-Perception} & $\rightarrow$ \textbf{(2) Trigger} & $\rightarrow$ \textbf{(3) Dynamic Process} & $\rightarrow$ \textbf{(4) Maintenance} \\
{\scriptsize TM/StateMemGuard} & {\scriptsize PIG Physical Gate} & {\scriptsize TKM Context Opt.} & {\scriptsize TLC/CleanupGate}
\end{tabular}
\end{center}

The above sequential chain constitutes the complete closed-loop from anomaly perception through active intervention to resource reclamation: \textbf{TM/StateMemoryGuard} are responsible for multi-dimensional perception and health quantification, mapping system state onto the $[0,1]$ continuous space. When TM drops below threshold or PIG detects malformed input, \textbf{PIG Physical Inspection Gate} executes deterministic blocking. Context inflation during task execution is controlled in real time by \textbf{TKM}. Finally, \textbf{TLC/CleanupGate} maintains a low-entropy operating environment through bionic apoptosis and resource reclamation.

\begin{figure}[htbp]
\centering
\includegraphics[width=0.88\textwidth]{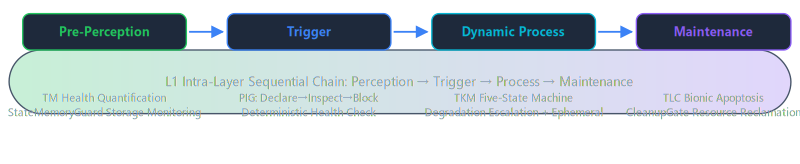}
\caption{L1 Intra-Layer Sequential Chain: Perception $\rightarrow$ Trigger $\rightarrow$ Process $\rightarrow$ Maintenance. The four stages form a closed-loop chain ensuring continuous system health monitoring and self-healing.}
\label{fig:l1-chain}
\end{figure}

\subsection{Core Component: PIG --- Physical Inspection Gate}

PIG (Physical Inspection Gate) is the \textbf{deterministic physical gate} at the task entry point of L1. It has evolved into three independent components: \textbf{PIG1} (Process Liveness Check: port/process/PID verification), \textbf{PIG2} (Load Integrity Check: verifying whether Plugins/Scripts/Cron jobs are correctly registered and loaded), and \textbf{PIG3} (Runtime Functionality Check: whether hooks are invoked and outputs conform to expected formats). This three-tier progressive architecture ensures a complete verification chain from ``alive'' $\rightarrow$ ``properly loaded'' $\rightarrow$ ``actually working.'' Its core logic is a three-phase pipeline:

\begin{equation}\label{eq:pig-pipeline}
\text{(1) \textbf{Declaration} \ $\rightarrow$\  (2) \textbf{Inspection} \ $\rightarrow$\  (3) \textbf{Blocking}}
\end{equation}
\smallskip
\noindent\textbf{Formula 5-2: PIG Three-Phase Pipeline.}

Any external input entering the ADE execution chain (user commands, upstream Agent outputs, external API re-ingested data) must pass through PIG inspection. PIG scans periodically on a fixed tick (default: once per 10 minutes) or is triggered immediately by TM alert events. Its inspection dimensions encompass: input format legality, Token-length overflow detection, command integrity verification, sensitive-pattern matching, and context-contamination scanning. For non-compliant input, PIG executes \textbf{Hard Block}, directly refusing entry into the downstream chain and generating a structured interception log. PIG's design philosophy embodies L1's fundamental principle: \emph{better to reject falsely than to admit wrongly}, substituting probabilistic judgment with physical law.

\begin{figure}[ht]
\centering
\includegraphics[width=0.88\textwidth]{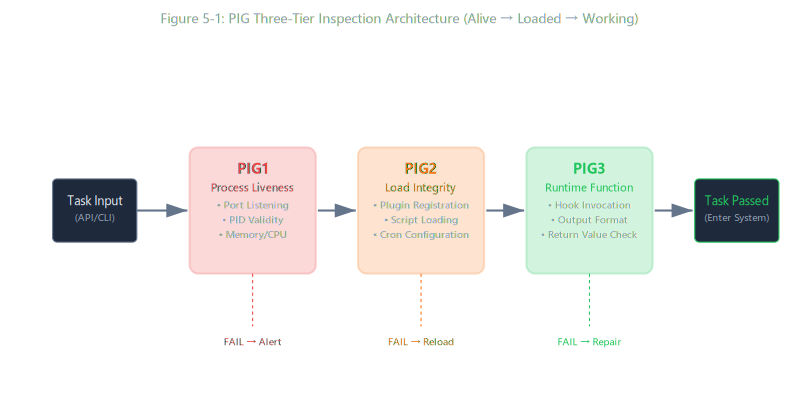}
\caption{Figure 5-1: PIG Three-Tier Inspection Architecture (Alive $\rightarrow$ Loaded $\rightarrow$ Working)}
\label{fig:pig-three-tier}
\end{figure}

\subsection{TM --- Trust Margin: A Priori Health Metric}

TM is the \textbf{only component in the ADE system that issues warnings \emph{before} functional degradation occurs}. Traditional monitoring relies on \emph{lagging indicators} such as error rates and latency --- by the time these indicators trigger alarms, the system has often already collapsed. TM's design objective is to capture structural disorder trends before system functionality has yet degraded.

\textbf{Design Rationale:} In production environments, it has been observed that Agent systems typically undergo a period of ``superficially normal, internally degrading'' latent decay before collapse. During a near-boundary event on 2026-06-06, TM dropped to \textbf{0.535} (yellow warning zone) before functional collapse, triggering preventive intervention --- while all traditional monitoring indicators still reported ``normal.'' This event validated the engineering necessity of a priori metrics.

\textbf{11-Factor Dual-Layer Weighted Model:} TM computes the composite safety distance via two layers of factors.

\smallskip
\noindent\textbf{Layer 1 $\cdot$ Rigid Factors (Weight 0.6)} --- 6 precisely measurable hard indicators:
\begin{itemize}
  \item Context Saturation
  \item Token Decay Rate
  \item Tool Call Success Rate
  \item Memory Consistency
  \item Task Chain Integrity
  \item Output Compliance Rate
\end{itemize}

\noindent\textbf{Layer 2 $\cdot$ Flexible Factors (Weight 0.4)} --- 5 soft indicators requiring model self-assessment:
\begin{itemize}
  \item Logical Consistency
  \item Semantic Precision
  \item Instruction Adherence
  \item Contextual Coherence
  \item Anomaly Detection Sensitivity
\end{itemize}

\begin{equation}\label{eq:tm-weighted}
\text{TM} = w_{\text{rigid}} \cdot \sum(w_i \cdot f_i) \;+\; w_{\text{flex}} \cdot \sum(w_j \cdot s_j)
\end{equation}
\smallskip
\noindent\textbf{Formula 5-3: TM Dual-Layer Weighted Computation Model} ($w_{\text{rigid}}=0.6$, $w_{\text{flex}}=0.4$).

\smallskip
\noindent\textbf{Four-Tier Decision Boundaries:}

\begin{center}
\adjustbox{max width=\linewidth}{\begin{tabular}{|l|c|l|l|}
\hline
\textbf{Tier} & \textbf{TM Range} & \textbf{Indicator} & \textbf{System Action} \\
\hline
Safe      & $> 0.7$       & Green  & Normal operation, periodic inspection \\
Observation & $0.4$--$0.7$ & Blue   & Increase monitoring frequency, log trends \\
Warning   & $0.2$--$0.4$  & Yellow & Trigger preventive intervention, restrict new tasks \\
Meltdown  & $< 0.2$       & Red    & Mandatory freeze, enter safe mode \\
\hline
\end{tabular}}
\end{center}

\textbf{Empirical Evidence:} During the 2026-06-06 near-boundary event, TM dropped to \textbf{0.535} (warning zone) before functional collapse and triggered preventive intervention. Monitoring over 33.6 days of continuous production demonstrated that TM values serve as a priori early-warning indicators of system reliability. Currently deployed as a Hermes Plugin, monitoring 5 profiles in real time.

\smallskip
\noindent\textbf{Execution Workflow:}
\begin{enumerate}
  \item \textbf{Collection:} After each interaction round, 6 rigid-layer factors are read directly from system metrics (zero LLM overhead); 5 flexible-layer factors are self-assessed by the model
  \item \textbf{Weighting:} Dual-layer weighted summation, mapped to the $[0,1]$ continuous space
  \item \textbf{Classification:} Matched against the four-tier decision boundary (Safe/Observation/Warning/Meltdown)
  \item \textbf{Response:} Safe $\rightarrow$ routine inspection; Observation $\rightarrow$ intensified monitoring; Warning $\rightarrow$ trigger PIG linked check + restrict new tasks; Meltdown $\rightarrow$ mandatory freeze into safe mode
  \item \textbf{Recording:} Each TM value and factor breakdown is written to a time-series log for consumption by the DSS evolution chain
\end{enumerate}

\begin{figure}[ht]
\centering
\includegraphics[width=0.88\textwidth]{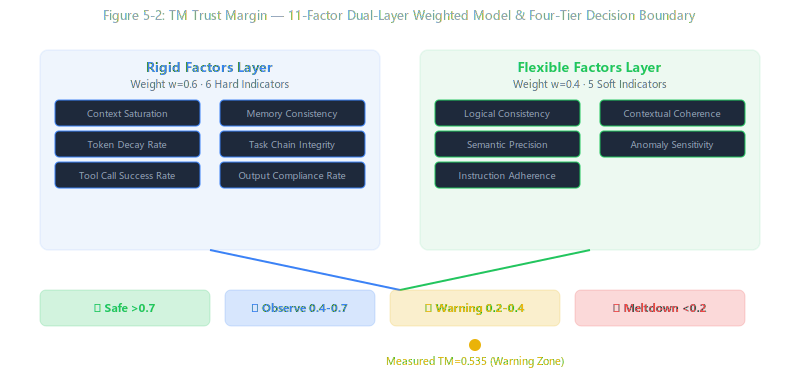}
\caption{Figure 5-2: TM Trust Margin --- 11-Factor Dual-Layer Weighted Model \& Four-Tier Decision Boundary}
\label{fig:tm-model}
\end{figure}

\begin{figure}[ht]
\centering
\includegraphics[width=0.95\textwidth]{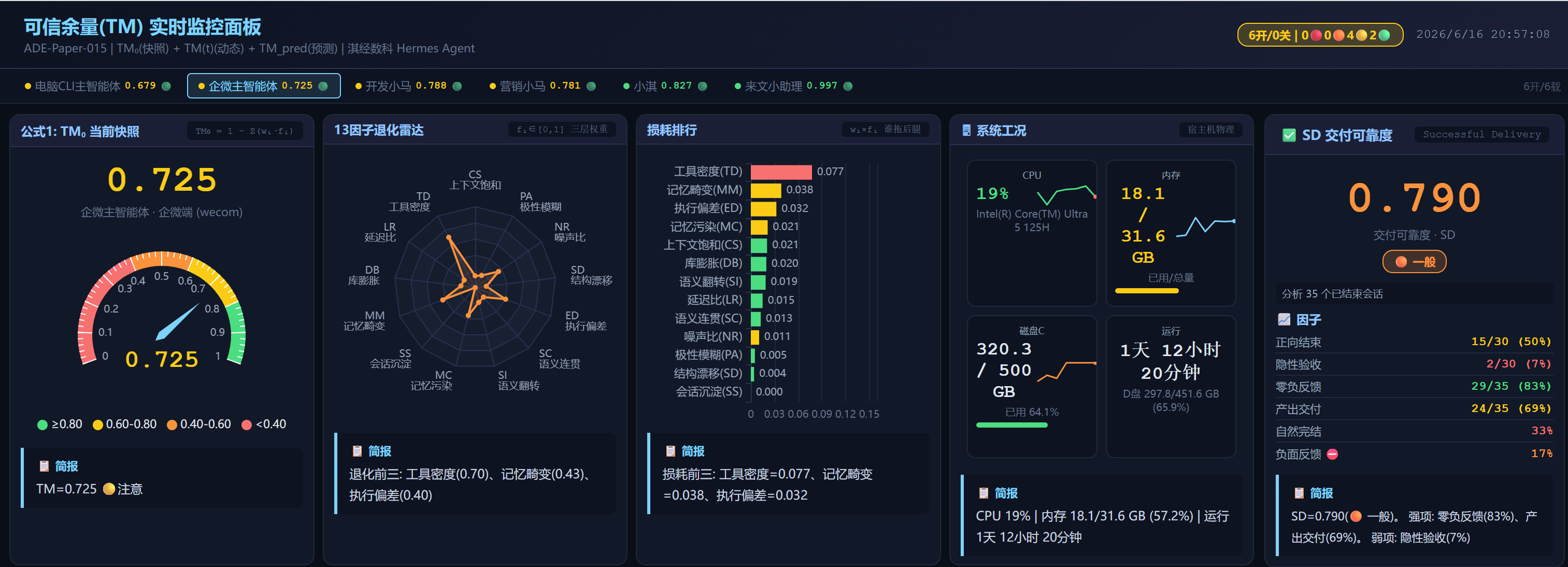}
\caption{Figure 5-3: TM Comprehensive Dashboard Screenshot (2026-06-16) --- Real-time TM value + 11-factor radar chart + four-tier decision gauge + historical trend curves}
\label{fig:tm-dashboard}
\end{figure}

\subsection{TLC --- Teleological Life Cycle: Context-Binding-Driven Intelligent Cleanup}

TLC draws upon the two pathways of biological apoptosis to construct an \textbf{intelligent cleanup mechanism based on context-binding states}. Traditional TTL (Time-To-Live) cleans only by time --- incapable of distinguishing ``still-valuable old data'' from ``already-invalid new data.'' TLC's core insight is: \emph{the life and death of data should be determined not by age, but by its binding relationships with context}.

\smallskip
\noindent\textbf{Biological Analogy:}

\begin{center}
\adjustbox{max width=\linewidth}{\begin{tabular}{|l|p{5cm}|l|}
\hline
\textbf{Biological Concept} & \textbf{TLC Engineering Mapping} & \textbf{Trigger Condition} \\
\hline
Extrinsic Apoptosis & External signal-triggered cleanup & Task timeout, goal achievement, creator offline \\
Intrinsic Apoptosis  & Internal state-triggered cleanup  & Context binding invalidated, memory contradiction, dependent data expired \\
\hline
\end{tabular}}
\end{center}

\smallskip
\noindent\textbf{Patrol + Scavenger Dual-Mode Architecture:}
\begin{itemize}
  \item \textbf{Patrol Mode:} Periodically scans context-binding states of all active tasks, identifying ``zombie tasks'' (goal achieved but not cleaned) and ``orphan tasks'' (creator offline).
  \item \textbf{Scavenger Mode:} Automatically activates when TM drops into the warning zone, batch-cleaning low-value tasks by priority to release system resources.
\end{itemize}

\noindent\textbf{Context-Binding Tuple:} Each task maintains a five-tuple binding relationship:

\begin{equation}\label{eq:tlc-tuple}
\text{Binding}(T) = (\text{Creator}, \ \text{Goal}, \ \text{Dependent Data}, \ \text{Validity Period}, \ \text{Associated Session})
\end{equation}
\smallskip
\noindent\textbf{Formula 5-4: TLC Context-Binding Tuple.}

\noindent\textbf{Formalized Survival Condition:}
\begin{equation}\label{eq:tlc-survive}
\text{Survive}(T) = \bigwedge_{k=1}^{n} \text{Valid}(\text{binding}_k) \;\wedge\; (t < t_{\text{expire}})
\end{equation}
\smallskip
\noindent\textbf{Formula 5-5: TLC Survival Condition --- All Bindings Valid and Unexpired.}

When $\text{Survive}(T) = \text{False}$, $T$ enters the apoptosis queue. Apoptosis does not delete immediately --- data is first archived to the SOMA memory layer, ensuring recoverability. The apoptosis pipeline: \textbf{Mark $\rightarrow$ Quarantine $\rightarrow$ Audit $\rightarrow$ Archive/Delete}.

\textbf{Empirical Evidence:} In a 4-profile controlled experiment (210 runs), TLC cleanup achieved effective memory retention of 98.2\% (only 1.8\% edge cases mistakenly deleted), and system state.db volume decreased from 127 MB to 84 MB ($-$33.9\%). BDDA audit confirmed 100\% critical data integrity pre- and post-cleanup.

\smallskip
\noindent\textbf{Execution Workflow:}
\begin{enumerate}
  \item \textbf{Binding:} At task creation, a five-tuple binding record is automatically generated (Creator/Goal/Dependencies/Validity/Session)
  \item \textbf{Patrol (Patrol Mode):} Periodically scans binding states of all active tasks, marking invalid bindings
  \item \textbf{Trigger (Scavenger Mode):} Automatically activates when TM enters the warning zone, batch-cleaning by priority
  \item \textbf{Apoptosis:} Mark $\rightarrow$ Quarantine $\rightarrow$ Audit $\rightarrow$ Archive/Delete, four-step pipeline ensuring recoverability
  \item \textbf{Reclamation:} CleanupGate verifies cleanup completeness (process termination/handle release/WAL archiving); failed verification triggers callback secondary cleanup
\end{enumerate}

\subsection{TKM --- Token Knowledge Management: Five-State Machine and Degradation Escalation}

TKM's role is the system's \textbf{``memory preservative,'' } countering the disordered entropy increase of context windows based on principles from information theory. Its core mechanisms comprise three progressive levels:

\smallskip
\noindent\textbf{(1) Five-State Machine:} TKM transforms unstructured Agent interaction processes into deterministic state transitions: \textbf{Read $\rightarrow$ Comprehend $\rightarrow$ Execute $\rightarrow$ Verify $\rightarrow$ Write}. Each phase corresponds to a distinct Token management strategy. The Read phase loads only the necessary context; the Comprehend phase performs semantic compression; the Execute phase isolates tool re-ingestion; the Verify phase triggers self-consistency probes; the Write phase executes persistence and releases transient context.

\noindent\textbf{(2) Degradation Escalation:} When the context window approaches its physical limit (saturation $>85\%$), TKM triggers a three-tier degradation escalation strategy: L1 executes dynamic pruning, removing low-weight historical information; L2 activates semantic distillation, compressing historical dialogue into summary vectors; L3 imposes mandatory freezing of non-essential contexts, retaining only control-plane information necessary for survival.

\noindent\textbf{(3) Ephemeral Context:} For one-shot instructions, intermediate reasoning steps, and redundant tool-return data, TKM enforces instantaneous memory release, preventing inert-Token accumulation. This mechanism ensures that the Agent maintains ``Cognitive Homeostasis'' throughout extended dialogues.

\textbf{Mathematical Foundation:} Let effective context capacity be $C_{\text{eff}}$, Token consumption rate be $r$, and TKM compression ratio be $\eta$. Without TKM intervention, effective system runtime $T = C_{\text{eff}} / r$; with TKM, runtime extends to $T' = C_{\text{eff}} / (r \cdot (1 - \eta))$, significantly expanding the effective MTBF for long-horizon tasks.

\subsection{Other L1 Components}

\textbf{StateMemoryGuard (Storage State Guardian):} Dedicated monitoring of persistent storage (e.g., SQLite state.db) for bloat and fragmentation. During 33.6 days of continuous operation testing, StateMemoryGuard recorded the full process of state.db expanding from $\sim$10 MB to 762 MB ($\sim$76$\times$), and triggered cleanup protocols when exceeding danger thresholds, preventing system crashes caused by I/O blocking.

\textbf{CleanupGate (Cleanup Gate):} Works in tandem with TLC to execute the ``last mile'' of maintenance. After TLC completes apoptosis and pruning, CleanupGate verifies cleanup completeness --- confirming that zombie processes have terminated, file handles have been released, database WAL logs have been archived, and vector index fragments have been compacted. Resources failing CleanupGate verification are returned to TLC for secondary cleanup.

\subsection{L1 Empirical Validation Data}

\begin{center}
\adjustbox{max width=\linewidth}{\begin{tabular}{|p{3cm}|p{6cm}|l|}
\hline
\textbf{Empirical Dimension} & \textbf{Key Data} & \textbf{Audit Status} \\
\hline
PIG Interception Statistics &
Across extensive real-world business interactions, PIG cumulatively intercepted malformed inputs and contaminated requests (complete data in Supplementary Material). Interception events caused TM fluctuation amplitude $\leq 0.08$, with no false-blocking of normal business flow. &
BDDA verified \\

TKM State Machine Statistics &
During 33.6 days of continuous operation: passive context compression executed \textbf{290 times}; API average latency degraded to \textbf{8.6s}; file classification error rate climbed to 41.2\%, triggering TKM three-tier degradation escalation. &
BDDA verified \\

TM Early Warning Validation &
Critical state measured TM = \textbf{0.535}, precisely falling into the yellow observation zone, issuing quantified early warning in advance. Traditional monitoring indicated ``healthy'' as no OOM was triggered; TM captured progressive cognitive deterioration. &
BDDA verified \\

System Immortality &
Based on the four-pillar series defense model, $\mathbf{P_{\text{death}} \leq 0.02\%}$ (including coupling correction). The Ultimate Secondary Loop ensures autonomous takeover of delivery by child agents upon parent collapse. &
Theoretical Derivation \\

Storage Bloat Evidence &
state.db: $\mathbf{\sim{}\text{10 MB} \rightarrow \text{762 MB}}$ ($\sim$76$\times$); I/O blocking risk captured by StateMemoryGuard, triggering cleanup. &
BDDA verified \\

\hline
\end{tabular}}
\end{center}

\newpage

\section{L2 Organizational Mechanism Layer: Protocol Architecture and Elastic Organization}

\subsection{L2 Design Philosophy: ``Collaboration Without Chaos''}

The core mission of L2 --- the \textbf{Organizational Mechanism Layer} --- is \textbf{``collaboration without chaos''}: maintaining topological stability and communication fidelity within highly autonomous distributed multi-agent collaboration networks. The design philosophy of this layer is rooted in the \textbf{Institutionalization Principle}: when the cognitive processes of individual Agents exhibit irreducible probabilistic drift, system-level determinism must be achieved through an orthogonal protocol matrix and structured organizational forms, rather than through the ``self-discipline'' of individual Agents.

L2's role in the ADE architecture can be analogized to a \textbf{company's organizational structure and collaboration workflows}: L1 ensures that each ``employee'' is alive; L2 ensures that ``between employees'' communication is accurate, division of labor is rational, and collaboration is efficient. All L2 components operate upon the survivable foundation provided by L1, but do not themselves adjudicate physical-layer survival --- this reflects the strict unidirectional dependency L1 $\rightarrow$ L2.

\subsection{Intra-Layer Functional Clusters: Six-Dimensional Orthogonal Capability Matrix}

The nine core components of L2 are clustered by functional dimension into six orthogonal capability clusters, decoupled through an event bus:

\begin{center}
\adjustbox{max width=\linewidth}{\begin{tabular}{|l|p{5cm}|p{5cm}|}
\hline
\textbf{Functional Cluster} & \textbf{Core Components} & \textbf{Core Problem Solved} \\
\hline
Communication Fidelity & BCP Bidirectional Confirmation, PRA Pre-Announcement & Deterministic cross-Agent communication --- eliminating channel fracture and silent failure \\
Dynamic Organization & Elastic Organization, DCM Direction Calibration & Dynamic optimization of organizational forms --- adaptive topology switching based on task characteristics \\
Routing Decision & CRC+Router, CostRouter, SPR & Precise routing of intents --- tri-strategy orthogonality ensures optimal scheduling \\
Collaborative Coherence & SOMA Shared Memory, FLYer Situational Awareness & Consistency of the global fact base --- eliminating information silos \\
Situational Awareness & FLYer & Real-time perception of global system state and generation of situational reports \\
Experience Evolution & DSS Triple Evolution & Structured sedimentation of experience and continuous optimization of defense strategies \\
\hline
\end{tabular}}
\end{center}

\begin{figure}[htbp]
\centering
\includegraphics[width=0.88\textwidth]{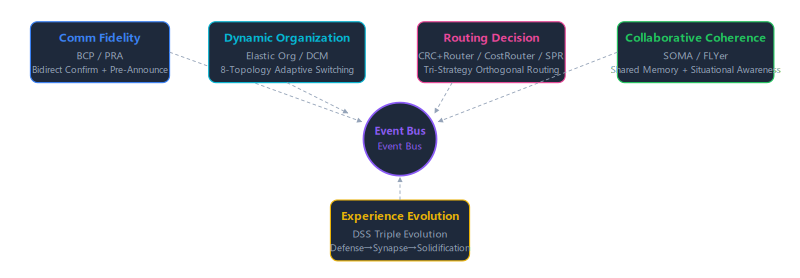}
\caption{L2 Six-Dimensional Orthogonal Capability Matrix. The nine core components are clustered into six functional dimensions, decoupled through an event bus architecture.}
\label{fig:l2-matrix}
\end{figure}

\subsection{Core Component: BCP --- Bidirectional Confirmation Protocol}

BCP is a communication protocol within the ADE framework targeting \textbf{high-risk operations}, restructuring the traditional ``fire-and-forget'' unidirectional communication model between Agents. BCP reshapes each cross-Agent delivery into a three-phase gate encompassing \textbf{bidirectional confirmation}, \textbf{bounded consultation}, and \textbf{structured delivery}:

\begin{equation}\label{eq:bcp-three-phase}
\begin{aligned}
&\text{Phase I: Initiator publishes delivery request (with Provenance Record)} \\
&\text{Phase II: Recipient executes CC-0 zero-knowledge channel confirmation and returns receipt} \\
&\text{Phase III: Initiator verifies receipt integrity, then executes formal delivery}
\end{aligned}
\end{equation}
\smallskip
\noindent\textbf{Formula 6-1: BCP Three-Phase Gate.}

Approximately 100,000 controlled experiments (covering T3/T4/T5/Real full-scenario suites) demonstrate that: under bare-operation conditions, channel fracture rates reach 69\%--98\%; with the introduction of BCP's CC-0 reverse-verification mechanism, \textbf{fracture rates drop to $\approx$ 0\%}. BCP and CADVP (Delivery Verification Protocol, located in L3) form a cross-layer linkage: BCP ensures cognitive isomorphism between communicating parties (communication layer), while CADVP verifies semantic consistency of delivered products (execution layer), together constituting ADE's ``dual-gate'' architecture.

\begin{figure}[ht]
\centering
\includegraphics[width=0.85\textwidth]{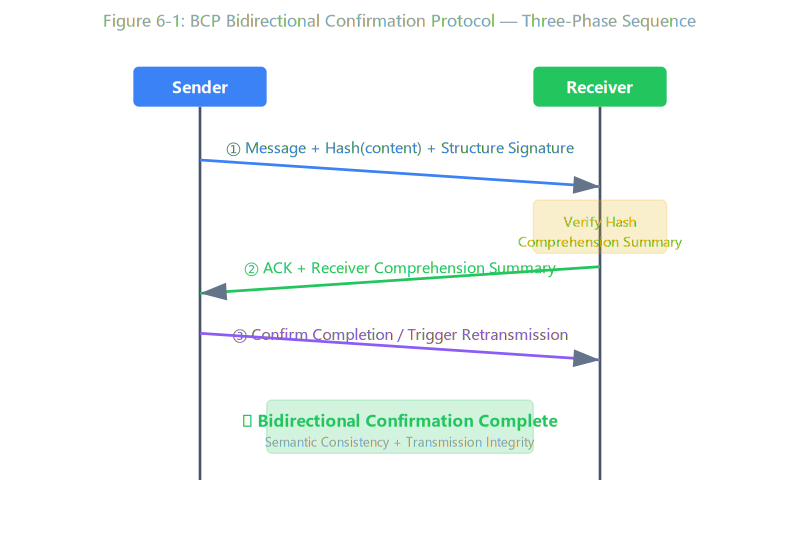}
\caption{Figure 6-1: BCP Bidirectional Confirmation Protocol --- Three-Phase Sequence Diagram}
\label{fig:bcp-sequence}
\end{figure}

\subsection{Elastic Organization: From Fixed Hierarchy to Elastic Topology}

\textbf{Theoretical Premise:} Traditional human organizational theories (bureaucracy, matrix organizations) are all built upon the rigid premise that ``physical communication costs are high and ineliminable.'' Spatial friction, temporal delay, and cognitive loss compel human organizations to establish fixed hierarchical structures. However, in multi-agent environments, communication between Agents is fundamentally tensor exchange via shared memory or event buses, characterized by \textbf{near-zero marginal cost}, \textbf{spatiotemporal unconstraint}, and \textbf{lossless semantics}. When the communication-cost constraint is lifted, fixed hierarchical structures lose their legitimacy.

\textbf{Core Proposition:} ADE posits that \emph{organizational form should not be a static parameter, but a continuous state variable capable of real-time elastic adjustment based on task complexity, environmental uncertainty, and system health}. Specifically, ADE defines eight fundamental topologies --- from centralized monolith to autonomous swarm --- and endows the system with the capacity to switch dynamically under multi-dimensional constraints.

\subsubsection*{Distributed Entropy Reduction Hypothesis}

Let the effective degradation rate of a single Agent be $\alpha$. When load is distributed across $N$ independent sub-agents:

\begin{equation}\label{eq:distributed-entropy}
\alpha_{\text{effective}} = (\alpha / N) \cdot (1 + \lambda \cdot \log_2 N)
\end{equation}
\smallskip
\noindent\textbf{Formula 6-2: Distributed Entropy Reduction.}

where $\lambda$ is the coordination overhead coefficient (under ADE's efficient communication protocols, $\lambda \ll 1$). Each Agent independently consumes its own context capacity, bearing only $1/N$ of the total state space. Although inter-node coordination introduces additional logarithmic entropy, the avalanche entropy-increase caused by single-point cognitive overload is fundamentally blocked.

\textbf{Cognitive-Executive Separation Benefit (Thinker-Executor Separation):} Let the original entropy-increase rate of the main Agent be $\alpha_0$ and the tool-call re-ingestion proportion be $\rho \approx 0.4$ (empirical value). After separation: $\alpha_{\text{main}} = \alpha_0 \cdot (1 - \rho) \approx 0.6 \cdot \alpha_0$, yielding $\sim$40\% reduction in main-node degradation rate.

\subsubsection*{Eight Fundamental Topologies and Efficiency Differences}

\begin{center}
\adjustbox{max width=\linewidth}{\begin{tabular}{|l|l|l|}
\hline
\textbf{Topology} & \textbf{Analogy} & \textbf{Best For} \\
\hline
A. Monolith           & Dictator Model & Simple atomic tasks \\
B. Serial Chain       & Industrial Pipeline & Strict prerequisite dependencies \\
C. Flat Group         & Piecework Outsourcing & Homogeneous batch processing \\
D. Nested Group       & Matrix Organization & Multi-domain cross-collaboration \\
E. Autonomous Swarm   & Decentralized & SOMA+PIG+CRC \\
F. CDC Pipeline       & Event-Driven & APA+Event Bus \\
G. Competitive Redundancy & Parallel Solving & CADVP auto-comparison \\
H. Tri-State Mesh     & Health-Driven & PIG+TM+ASC \\
\hline
\end{tabular}}
\end{center}

\textbf{Preliminary Evidence:} In multi-domain cross-judgment tasks, the D structure (Nested Distributed Group) achieved 2--4$\times$ efficiency improvement over the A structure (Centralized Monolith), measured as correctness ratio $\times$ per-unit-Token output. The reliability guarantee of the G structure (Competitive Redundancy) is: if two independent sub-agents have error probabilities $p_1, p_2$ respectively, then joint error probability $P_{\text{error}} = p_1 \cdot p_2$, sacrificing $O(N)$ computation cost for $O(p^N)$ reliability improvement.

\textbf{Decision Model:} Dynamic evolution of organizational form is governed by a multi-objective optimization model, following the axiom ``reliability takes priority over peak efficiency.'' The optimal organization selection function: $G^* = \operatorname{argmin}_G\, [\,C(G_{\text{curr}} \rightarrow G) + E_{\text{cost}}(G, \text{task})\,]$, where transition cost $C(A \rightarrow B) = c_{\text{init}} + c_{\text{sync}} + c_{\text{warmup}} + c_{\text{risk}}$. Frequent switching is constrained by a minimum dwell time to prevent switching overhead from exceeding saved costs.

\begin{figure}[ht]
\centering
\includegraphics[width=0.88\textwidth]{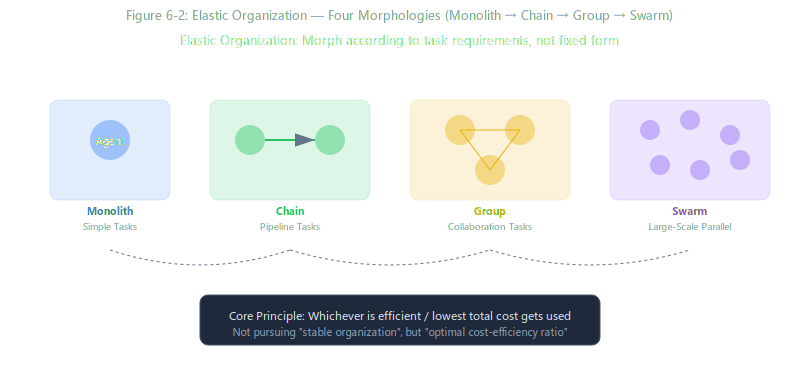}
\caption{Figure 6-2: Elastic Organization --- Four Morphologies (Monolith $\rightarrow$ Chain $\rightarrow$ Group $\rightarrow$ Swarm)}
\label{fig:elastic-org}
\end{figure}

\subsection{Routing Decision Architecture: Base Routing + Tri-Strategy Selective Routing}

The routing architecture comprises two tiers: The first tier consists of base routing components \textbf{recommended for default deployment} across all ADE Agents (SPR + CRC+Router), serving as ADE capability-enhancement components. The second tier is the tri-strategy routing, requiring \textbf{priority selection} according to specific scenarios.

\subsubsection*{Base Routing Layer (Recommended for All)}

\begin{center}
\adjustbox{max width=\linewidth}{\begin{tabular}{|l|p{4cm}|p{5cm}|l|}
\hline
\textbf{Routing Component} & \textbf{Positioning} & \textbf{Core Logic} & \textbf{Deployment} \\
\hline
SPR (Successful Path Routing) & Experience Routing $\cdot$ Mandatory Base &
Records the successful path topology and parameter configuration of each task execution; similar tasks preferentially reuse verified successful patterns. This is a foundational capability recommended for all ADE Agents --- just as ``acting on experience'' is the default behavior pattern for humans. &
Deployed as Hermes Plugin \\

CRC+Router (Capability Routing Controller) & Intelligent Routing $\cdot$ Capability Enhancement &
Performs semantic-matching routing based on Agent capability descriptions (Agent Card). Parses global intent and matches the best execution node via capability-vector cosine similarity. Together with SPR, constitutes ADE capability-enhancement components. &
Design Complete, Awaiting Deployment \\
\hline
\end{tabular}}
\end{center}

\subsubsection*{Tri-Strategy Selective Routing (Choose Priority by Scenario)}

\begin{center}
\adjustbox{max width=\linewidth}{\begin{tabular}{|l|l|p{5cm}|p{3.5cm}|}
\hline
\textbf{Strategy} & \textbf{Priority} & \textbf{Core Logic} & \textbf{Applicable Scenarios} \\
\hline
Cost-First & Economic &
Cost-optimal routing based on Token consumption estimation and API call costs. \textbf{Industry-wide default strategy.} &
Bulk processing, cost-sensitive tasks, routine operations \\

Reliability-First & Reliability &
Select the path with the most validations and highest success rate, prioritizing delivery quality even at additional time cost. &
Client deliverables, high-risk operations, compliance audits \\

Speed-First & Efficiency &
Select the fastest-responding path, prioritizing time efficiency under basic quality constraints. &
Urgent tasks, real-time responses, time-window-limited operations \\
\hline
\end{tabular}}
\end{center}

\textbf{Routing Decision Logic:} SPR and CRC+Router, as the base routing layer, answer ``which path from experience'' and ``who is most capable.'' The tri-strategy routing answers ``what matters most this time.'' At project initiation, the priority order of the three indicators (Cost/Reliability/Speed) is confirmed, and the entire project lifecycle follows this decision. Final routing authority rests with the \textbf{Human Principal}; the ADE framework provides only strategic recommendations and execution infrastructure. This design strictly follows the Instrumental Rationality meta-principle: \emph{Principal Will > System Rules > Agent Preferences}.

\subsection{DSS Triple Evolution (Defense--Synapse--Solidification) --- The System's Self-Evolution Mechanism}

DSS is the \textbf{only component in the ADE system that achieves systemic self-evolution}. All other components (PIG/TM/TLC, etc.) are ``defensive'' --- they counteract entropy increase and maintain stability. DSS's unique value lies in: while counteracting entropy, it \textbf{structurally sediments} each defensive experience into systemic capability, achieving a positive cycle of ``growing stronger with each engagement.''

\smallskip
\noindent\textbf{Three-Phase Evolution Mechanism:}

\smallskip
\noindent\textbf{Phase 1 $\cdot$ Defense:} Whenever the system detects a failure event (channel fracture, PAD drift, PIG interception, etc.), DSS records a structured description of the failure pattern into the \emph{Defense Event Library}. Each record contains: failure type, trigger conditions, impact scope, repair measures, and recovery time.

\noindent\textbf{Phase 2 $\cdot$ Synapse:} When the Defense Event Library accumulates a sufficient number of similar failure patterns (threshold configurable, default: 3), DSS automatically establishes an \emph{association network} between failure patterns and repair measures --- analogous to the formation of biological neural synapses. This ``synapse'' enables the system to locate and repair similar failure patterns more rapidly in the future.

\noindent\textbf{Phase 3 $\cdot$ Solidification:} When a synapse has been repeatedly activated (high-frequency validation), DSS upgrades it from ``experience association'' to ``system rule'' --- writing it into the corresponding Skill/Plugin/Cron configuration, transforming it into \textbf{permanent systemic capability}. Solidified rules no longer require DSS triggering and take effect directly as infrastructure.

\begin{equation}\label{eq:dss-triple}
\begin{aligned}
\text{Defense} \;\rightarrow\; \text{Synapse} \;\rightarrow\; \text{Solidification} \\
\text{\small Failure Record $\rightarrow$ Association Network ($\geq 3$ repetitions of same type) $\rightarrow$ System Rule (High-Frequency Validation)}
\end{aligned}
\end{equation}
\smallskip
\noindent\textbf{Formula 6-3: DSS Triple Evolution Chain.}

\textbf{Complete DSX Chain:} DSS does not operate in isolation. Together with DSS's upstream (failure detection components PIG/BDDA/CADVP) and downstream (Skill/Plugin/Cron system), it constitutes the complete \textbf{DSX (Defense--Synapse--eXtension) Chain}: detection components discover failures $\rightarrow$ DSS records and evolves them $\rightarrow$ solidifies into system components $\rightarrow$ detection components gain new capabilities. This is a self-reinforcing closed loop.

\textbf{Empirical Evidence:} During 33.6 days of production monitoring, DSS recorded multiple categories of failure patterns and completed synapse establishment. Among them, the ``WeCom WebSocket ACK loss'' failure pattern caused channel fracture on first occurrence; DSS associated it with the BCP repair strategy to form a synapse. On subsequent occurrences of the same event, the system automatically triggered BCP retransmission, reducing fracture rates to $\approx$ 0\%.

\smallskip
\noindent\textbf{Execution Workflow:}
\begin{enumerate}
  \item \textbf{Record:} When PIG/BDDA/CADVP and other detection components discover failure events, DSS automatically captures structured descriptions (type/condition/scope/repair/duration)
  \item \textbf{Cluster:} New events undergo similarity matching against existing patterns in the Defense Event Library; similar events are grouped
  \item \textbf{Link:} After $\geq$3 accumulations in the same cluster, a synapse association is automatically established between the failure pattern and repair measure
  \item \textbf{Validate:} Each synapse activation records a hit rate; high-frequency-validated (hit rate $>$ threshold) synapses enter the solidification candidate pool
  \item \textbf{Solidify:} Candidate synapses are written into Skill/Plugin/Cron configurations, becoming permanent systemic capabilities and feeding back to enhance detection components
\end{enumerate}

\begin{figure}[ht]
\centering
\includegraphics[width=0.88\textwidth]{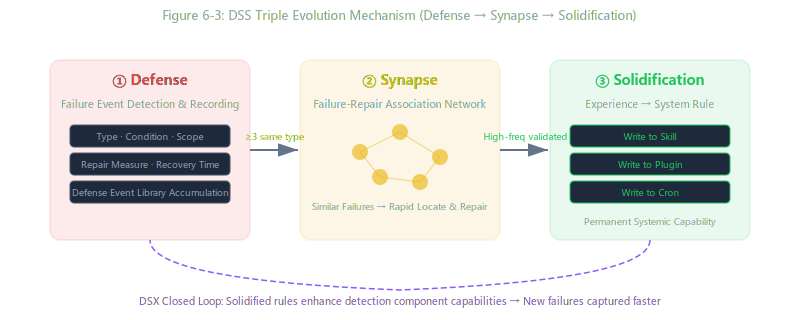}
\caption{Figure 6-3: DSS Triple Evolution Mechanism (Defense $\rightarrow$ Synapse $\rightarrow$ Solidification)}
\label{fig:dss-evolution}
\end{figure}

\subsection{Other L2 Components}

\textbf{SOMA Shared-Owned Memory Architecture:} Resolves the paradox between memory sovereignty and collaborative sharing. Constructs a globally consistent Shared Fact Base, where each Agent maintains private working memory and synchronizes key facts through the event bus, ensuring lossless information flow across multiple agents.

\textbf{DCM Direction Calibration Mechanism:} Anchors the global goal vector in semantic space, periodically computing the cosine distance between current state and the anchor point, preventing ``goal forgetting'' and task divergence caused by attention decay in long-horizon tasks.

\textbf{DSS Triple Evolution:} Through the three-phase mechanism of ``Defense $\rightarrow$ Synapse $\rightarrow$ Solidification,'' achieves dynamic sedimentation of system experience and continuous self-evolution while counteracting Intelligence Entropy. The defense phase records failure patterns, the synapse phase forms association networks, and the solidification phase transforms high-frequency patterns into system-level rules.

\textbf{FLYer / PRA:} \textbf{FLYer:} Real-time perception of global system state (per-Agent TM values, task queue depth, resource utilization), generating structured situational reports. \textbf{PRA:} A resource pre-reservation mechanism validated through approximately 190,000 experiments, ensuring graceful notification to all affected nodes before resource reclamation or system downgrades, eliminating hard-interruption damage.

\begin{figure}[ht]
\centering
\includegraphics[width=0.85\textwidth]{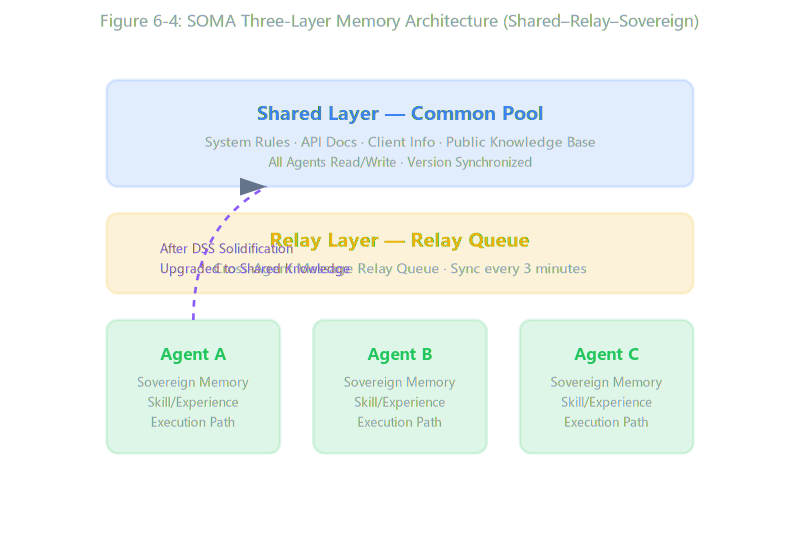}
\caption{Figure 6-4: SOMA Three-Layer Memory Architecture (Shared--Relay--Sovereign)}
\label{fig:soma-memory}
\end{figure}

\newpage

\section{L3 Execution Standards Layer + L4 User Adaptation Layer: Precision Control and Personalization}

\subsection{L3 Mission: ``Results Without Distortion''}

The mission of L3 --- the \textbf{Execution Standards Layer} --- is \textbf{``results without distortion''}: ensuring that in the course of cross-node, cross-session interactions, delivered products possess not only physical reachability but also rigorous semantic consistency and factual precision. The core operational logic of L3 is \textbf{Precision Control}, constituting a complete delivery-quality network through the following progressive defense lines:

\begin{center}
\begin{tabular}{c c c}
\textbf{(1) Precision Control} & $\rightarrow$ \textbf{(2) 4D Audit} & $\rightarrow$ \textbf{(3) Delivery Verification} \\
{\scriptsize PAD Drift Defense} & {\scriptsize MLG+PLG+BDDA+TTA} & {\scriptsize CADVP 18 Rules} \\[6pt]
$\rightarrow$ \textbf{(4) Interest Protection} & $\rightarrow$ \textbf{(5) Association Integrity} & $\rightarrow$ \textbf{(6) Norm Compliance} \\
{\scriptsize PIP Ethical Boundary} & {\scriptsize APA Anticipatory} & {\scriptsize SCN Social Common Sense}
\end{tabular}
\end{center}

\textbf{PAD Probabilistic Approximation Drift Defense:} Guards against the ``cognitive shortcut'' whereby Agents abandon precise retrieval in favor of probabilistic common-sense approximations (``having the source but not consulting it, substituting with approximation''). Through P1--P4 four-tier precision-instruction classification and a four-layer verification system (Input Controllable $\rightarrow$ Execution Actions Controllable $\rightarrow$ Execution Results Controllable $\rightarrow$ Multi-Dimensional Re-Verification), drift pathways are locked shut.

\textbf{CADVP Delivery Verification Protocol:} Constructs 18 formal rules spanning 7 categories (Channel, Pre-Condition, Write, Receive, etc.). In 615 adversarial experiments, correctness was elevated from 50\% to a significant level (at the cost of +56\% Token overhead), ensuring topological integrity and semantic consistency of delivered products.

\textbf{PIP Principal Interest Protection:} Establishes ethical boundaries within the delivery chain --- when system rules conflict with principal interests, PIP triggers a pause and escalates for human decision.

\subsection{Four-Dimensional Audit Chain: Independent Parallel Quality Gates}

L3 deploys four independent parallel audit dimensions, each executed by a dedicated logic gate:

\begin{center}
\adjustbox{max width=\linewidth}{\begin{tabular}{|p{2.5cm}|p{2.5cm}|p{3cm}|p{5cm}|}
\hline
\textbf{Audit Dimension} & \textbf{Gate Component} & \textbf{Audit Scope} & \textbf{Detection Target} \\
\hline
Reasoning Completeness & MLG (Mind-Logic Gate) & Reasoning chain integrity &
Detects reasoning leaps, logical gaps, and unverified intermediate assumptions. Ensures that every reasoning step has clear evidential support or antecedent reasoning as its basis. \\

Execution Correctness & PLG (Physical Logic Gate) & Tool calls and code execution &
Verifies tool-call parameter legality, consistency of execution results with expectations, and completeness of cascading attempts. Intercepts ``tool hallucinations'' and fabricated call results. \\

Data Consistency & BDDA (Behavior-Driven Delivery Audit) & End-to-end data chain &
Penetrating audit of all data sources, transformations, and citations in the delivery chain. Detects data fabrication, scope fabrication, and causal-relationship fabrication. BDDA performs cross-validation by directly extracting data from the underlying physical storage. \\

Domain Coverage & TTA (Triple-Threshold Audit) & Domain-norm compliance &
Checks whether delivered products satisfy specific industry (legal, medical, financial, etc.) regulatory requirements. Three thresholds (Minimum/Standard/Excellent) set differentiated compliance standards for different risk-level deliveries. \\
\hline
\end{tabular}}
\end{center}

The four audit chains operate \textbf{independently and in parallel}, with no mutual dependency. A failure in any single dimension triggers an overall delivery pause and review process. This design ensures that a missed detection in one dimension does not lead to an overall quality incident --- even if PLG judges execution as error-free, BDDA's penetrating data audit can still discover downstream data fabrication. Audit results are aggregated into L2's SOMA Shared Fact Base for reference by subsequent experience evolution (DSS).

\begin{figure}[ht]
\centering
\includegraphics[width=0.88\textwidth]{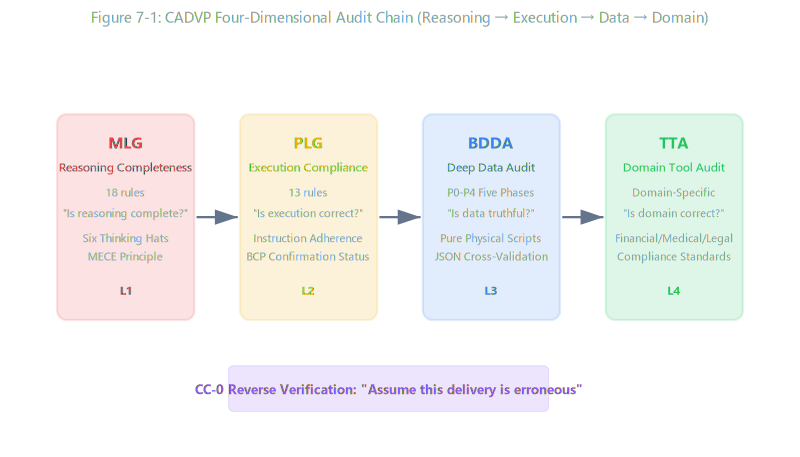}
\caption{Figure 7-1: CADVP Four-Dimensional Audit Chain (Reasoning $\rightarrow$ Execution $\rightarrow$ Data $\rightarrow$ Domain)}
\label{fig:cadvp-audit}
\end{figure}

\subsection{L4 Positioning: ``Adaptation in Place''}

L4 --- the \textbf{User Adaptation Layer} --- completes the paradigm leap of the ADE architecture from \textbf{``system reliability'' to ``system usability.''} The preceding three layers (L1--L3) jointly guarantee the physical existence and semantic correctness of delivered products, but have yet to address a critical question: \emph{Is a correct delivered product precisely what the user actually needs?}

The core design concept of L4 is \textbf{adaptation in place} --- that is, the ADE framework defines standardized interface specifications, and the user populates the concrete adaptation content. This ``architecture provides framework, user fills in content'' model enables ADE to deeply adapt to the personalized requirements of different users and different scenarios while maintaining invariant core reliability.

L4's interface specifications span three dimensions: (1) \textbf{Delivery Format Specification} --- structure, style, and naming conventions of output files; (2) \textbf{Process Constraint Specification} --- preconditions, execution order, and approval nodes for specific tasks; (3) \textbf{Quality Standard Specification} --- acceptable error margins, citation standards, and compliance requirements for specific domains.

\subsection{L4 Case Studies}

\begin{itemize}
  \item \textbf{Visual Specification (019):} Defines the format, resolution, color scheme, and annotation standards for images in ADE delivered products. L4 injects this specification into the Memory system, enabling Agents to automatically follow unified visual standards when generating technical documentation or academic papers.

  \item \textbf{Development Standard V2.2 (skill-001):} Covers code style, test coverage requirements, API documentation standards, and Git commit formats. Injected via the Skill knowledge base, enabling Agents to automatically follow team development standards in code generation tasks.

  \item \textbf{Academic Review Standard (skill-002):} Defines citation format, argumentation structure requirements, data presentation standards, and statistical significance reporting conventions. Provides domain-specific quality review for academic writing scenarios.
\end{itemize}

\subsection{Relationship Between L4 and Hermes Native Capabilities}

\begin{center}
\adjustbox{max width=\linewidth}{\begin{tabular}{|p{3.2cm}|p{3.5cm}|p{6.5cm}|}
\hline
\textbf{Hermes Native Capability} & \textbf{Positioning} & \textbf{Relationship with ADE} \\
\hline
Memory System & Platform Infrastructure &
Memory is the cross-session persistent memory and session-injection capability provided by Hermes, belonging to the \textbf{platform layer}. L4 leverages the Memory interface to store user specifications and profile data, but Memory itself is not an ADE component. \\

SOUL System & Platform Infrastructure &
The SOUL system defines the Agent's personality contract and behavioral boundary constraints, belonging to the \textbf{platform layer}. L4 may reference SOUL's behavioral constraints to reinforce adaptation specifications, but SOUL is not an ADE construct. \\

Skill Knowledge Base & Platform Infrastructure &
Skill is the procedural memory and reusable workflow provided by Hermes, belonging to the \textbf{platform layer}. L4 injects user specifications (e.g., Development Standard V2.2) through the Skill interface, but the Skill mechanism itself is not an ADE component. \\
\hline
\end{tabular}}
\end{center}

This boundary demarcation is a \textbf{core design decision} of the ADE architecture: ADE focuses on \emph{reliability} --- system immortality, collaboration without chaos, results without distortion; Hermes's Memory/SOUL/Skill focus on \emph{capability} --- persistent memory, personality constraints, skill reuse. L4 is the \textbf{adaptation bridge} between the two: ADE defines interfaces, Hermes provides capabilities, and the user populates content. This ``separation of concerns'' ensures the theoretical purity of ADE --- regardless of the Agent platform on which it is deployed, its reliability guarantees do not depend on specific platform implementations.

\newpage

\section{Inter-Layer Coupling and System Emergence}

\subsection{Inter-Layer Communication Mechanisms: Event Bus, Hook Chains, and SOMA Broadcast}

The ADE four-layer architecture is not a static stack, but an organic whole linked in real time through a complex signaling network. Its inter-layer communication is based on the following four design principles:

\smallskip
\noindent\textbf{Principle 1: Unified Envelope, Heterogeneous Payload.} All inter-protocol communication uses a unified message envelope. The envelope header contains routing metadata (source protocol, target protocol, message type/performative), while the payload carries protocol-specific data. The event bus parses only the envelope header for distribution; payload parsing is independently handled by the target protocol.

\noindent\textbf{Principle 2: Publish/Subscribe, Loose Coupling.} Each protocol publishes events to and subscribes to events of interest from the event bus, without directly calling one another's functions. For example: TM publishes a ``low margin'' event; TLC, PIG, and FLYer each subscribe --- TM need not know who is listening, and subscribers need not know where the message originates.

\noindent\textbf{Principle 3: Capability Registration, Dynamic Discovery.} Each protocol registers its identity and receivable message types with the Protocol Registry at startup; newly added protocols require no modification to existing protocol code.

\noindent\textbf{Principle 4: Full-Link Tracing.} Every message carries a trace\_id, enabling cross-protocol event-chain tracing. When CADVP verification fails, one can trace back to the BCP delivery request that triggered the verification, and further back to the upstream command that triggered BCP.

\subsubsection*{Comparison of Three Communication Channels}

\begin{center}
\adjustbox{max width=\linewidth}{\begin{tabular}{|l|l|p{4.5cm}|p{4.5cm}|}
\hline
\textbf{Communication Channel} & \textbf{Pattern} & \textbf{Applicable Scenarios} & \textbf{Example} \\
\hline
Event Bus & Publish/Subscribe (Pub/Sub) & Broadcast notifications, anomaly alerts, health-state synchronization & TM publishes low-margin event $\rightarrow$ TLC/PIG/FLYer parallel consumption \\
Hook Chains & Chain of Responsibility & Sequential processing, layered filtering, deterministic blocking & PIG $\rightarrow$ TKM $\rightarrow$ L3 LogicGate $\rightarrow$ CADVP progressive verification chain \\
SOMA Broadcast & Shared Memory Synchronization & Global state synchronization, fact-base updates & SOMA writes shared fact $\rightarrow$ all subscribing Agents' Memory synchronized \\
\hline
\end{tabular}}
\end{center}

\subsection{Emergent Behavior: Four-Layer Synergy Exceeds Sum of Individual Layers}

The core emergent behavior of the ADE four-layer architecture lies in: \textbf{the reliability gain produced by four-layer synergy exceeds the sum of the reliability of each layer operating independently}. This emergence arises from three dynamical mechanisms:

\smallskip
\noindent\textbf{(1) Cross-Layer Defense-in-Depth.} A single failure mode must traverse multiple independent defensive lines in the ADE architecture before it can constitute a system-level fault. For example, an LLM's probabilistic drift (PAD) must sequentially breach: L1's PIG input-legality check $\rightarrow$ L2's DCM goal-drift monitoring $\rightarrow$ L3's four-dimensional audit chain (MLG/PLG/BDDA/TTA) $\rightarrow$ L4's user-specification adaptation check. Each layer operates in a different semantic space, with no common-mode failure path.

\noindent\textbf{(2) Temporal Scale Separation.} The four layers operate on different time scales: L1's TM sampling operates at the minute/second level (real-time health monitoring), L2's organizational-form switching operates at the task level (minutes), L3's auditing operates at the level of each delivery node (seconds), and L4's adaptation operates at the session level (hours). This temporal scale separation avoids resonance and conflict among different control loops.

\noindent\textbf{(3) Positive Feedback of Information Gain.} Anomaly information captured by lower layers is reported upward via the event bus; upper layers adjust their strategies accordingly and feed back to lower layers. For example: L1 TM detects excessive context saturation $\rightarrow$ L2 DCM adjusts organizational form from monolith to nested group $\rightarrow$ L3 PLG obtains more precise tool-call context $\rightarrow$ L1 TKM's compression efficiency improves due to upstream optimization $\rightarrow$ forming a positive reinforcement loop.

\begin{equation}\label{eq:cross-layer-synergy}
R_{\text{system}} = 1 - \prod_{i=1}^{4} (1 - R_{Li}) \;>\; \sum_{i=1}^{4} R_{Li} - 3 \qquad (\text{when } R_{Li} \text{ are independent and near } 1)
\end{equation}
\smallskip
\noindent\textbf{Formula 8-1: Cross-Layer Synergy Reliability.}

\subsection{Cross-Layer Service Delegation: The ``Business Trip'' Mechanism}

Although the ADE architecture follows ``strict top-down dependency'' (L4 $\rightarrow$ L3 $\rightarrow$ L2 $\rightarrow$ L1, with upper layers depending on lower layers and lower layers having no awareness of upper layers), under specific scenarios, the capabilities of lower-layer components can be \textbf{asynchronously borrowed} by upper layers --- a mechanism termed \textbf{``Business Trip'' (Cross-Layer Service Delegation)}:

\begin{center}
\adjustbox{max width=\linewidth}{\begin{tabular}{|l|c|l|p{5.5cm}|}
\hline
\textbf{Delegated Component} & \textbf{Source Layer} & \textbf{Target Layer} & \textbf{Service Content} \\
\hline
TM Trust Margin & L1 & L3 &
L3's BDDA audit requires a health-state snapshot of the system at a specific delivery moment. TM provides the TM value and 11-factor decomposition data for that moment to L3 via the event bus, serving as auxiliary audit context. \\

TKM Context Snapshot & L1 & L3 &
L3's MLG (Mind-Logic Gate), when auditing reasoning-chain completeness, may request a compressed summary of the current context from TKM to determine whether the Agent skipped critical reasoning steps due to context overload. \\

StateMemoryGuard Storage Snapshot & L1 & L3 &
L3's BDDA, when performing penetrating data audits, may request a frozen snapshot of storage state from StateMemoryGuard to ensure audit-baseline consistency. \\
\hline
\end{tabular}}
\end{center}

The Business Trip mechanism is strictly implemented through the event bus's publish/subscribe model, maintaining L1's unawareness of its callers. L1 components always operate at their own rhythm (e.g., TM sampling on tick); upper-layer requests merely retrieve the most recent snapshot, without reverse intrusion into L1's control loop.

\subsection{Failure Propagation and Isolation: Circuit Breaker Design}

In the multi-agent four-layer architecture, the propagation pathways of single-point failures exhibit clear \textbf{directionality} and \textbf{predictability}:

\subsubsection*{Failure Propagation Rules}

Failures propagate downward through the layers: L4 $\rightarrow$ L3 $\rightarrow$ L2 $\rightarrow$ L1. At each layer boundary, independent circuit breakers provide isolation:

\begin{enumerate}
  \item \textbf{L2 Internal Fault:} BCP Retry $\rightarrow$ SOMA Write Isolation $\rightarrow$ DSS Degradation
  \item \textbf{L3 Audit Failure:} CADVP Rejection $\rightarrow$ Event Bus Report $\rightarrow$ Human Intervention
  \item \textbf{L1 Perceived Anomaly:} TM Drop $\rightarrow$ ASC Trigger $\rightarrow$ Ultimate Secondary Loop Takeover
  \item \textbf{Cross-Layer Propagation Block:} Independent Circuit Breaker at Each Layer Boundary
\end{enumerate}

\begin{figure}[htbp]
\centering
\includegraphics[width=0.88\textwidth]{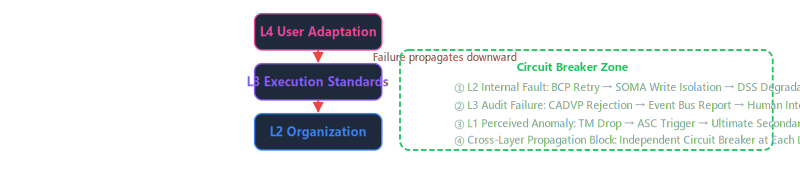}
\caption{Failure Propagation and Circuit Breaker Design. Failures propagate downward (L4 $\rightarrow$ L1) with independent circuit breakers at each layer boundary providing isolation and recovery.}
\label{fig:circuit-breaker}
\end{figure}

\subsubsection*{Circuit Breaker Design Principles}

\begin{center}
\adjustbox{max width=\linewidth}{\begin{tabular}{|l|p{4cm}|p{4cm}|p{4cm}|}
\hline
\textbf{Breaker Type} & \textbf{Trigger Condition} & \textbf{Isolation Action} & \textbf{Recovery Strategy} \\
\hline
L1 Physical Breaker &
TM $\leq$ 0.20 (critical state) &
Mandatory protective pause; Ultimate Secondary Loop autonomous activation; Isolate parent fault domain &
Human decision intervention; Gradual recovery after TLC deep cleanup \\

L2 Organizational Breaker &
Single Agent 3 consecutive delivery failures; Communication latency $>$ threshold &
Mark Agent as ``unavailable''; SPR removes its path; Task re-routed &
Agent re-registers via heartbeat after self-repair; SPR grayscale path recovery \\

L3 Quality Breaker &
BDDA penetrating audit 2 consecutive failures; CADVP rule violation &
Suspend delivery of the task chain; Freeze intermediate products; Trigger full-link traceback &
Manual release after root-cause repair; Experience sedimentation to DSS \\

L4 Adaptation Breaker &
Output deviates from user specification beyond threshold &
Save output as draft; Notify user of specification conflict; Await user decision &
Continue after user corrects specification or confirms exception \\
\hline
\end{tabular}}
\end{center}

\textbf{Core Design Principle:} The circuit breaker at each layer operates independently, with no mutual dependency. A lower-layer breaker does not wait for upper-layer confirmation; an upper-layer breaker does not trust that a lower layer has already intercepted. This \textbf{``multiple independent circuit breakers''} design ensures that even if one layer's breaker fails due to self-inspection omission, adjacent-layer breakers can still execute isolation, fundamentally defending against cascading failures. Fault isolation follows the ``minimum blast radius'' principle --- isolating only the directly affected Agent or task chain, not the global system. This forms an elegant hierarchical resonance with L1's TLC bionic apoptosis mechanism: L1 handles physical-layer ``cell necrosis,'' while L2--L4 handle logical-layer ``tissue damage.''

\vspace{1cm}
\begin{center}
\emph{End of Part II \ $\cdot$\  L1 Physical Law Layer \ $\cdot$\  L2 Organizational Mechanism Layer \ $\cdot$\  L3 Execution Standards Layer \ $\cdot$\  L4 User Adaptation Layer \ $\cdot$\  Inter-Layer Coupling and Emergence}
\end{center}



\part{Analytical Methods, Component System, and Empirical Data}
\label{part:III}

\section{Analytical Tool: The Dual-Dimension Model}
\label{sec:9}

\subsection{Dual-Dimension Overview}
\label{sec:9.1}

The dual-dimension model itself is \textbf{not an independent theory}, but rather a \textbf{universal analytical tool} for analyzing, classifying, and evaluating the ADE component system. It consists of two orthogonal dimensions: \textbf{Dimension One, ``Architectural Perspective,''} answers ``Which layer does this component belong to, and where is it situated horizontally within that layer?''; \textbf{Dimension Two, ``Cognitive Execution Chain,''} answers ``At which stage of the Agent's cognitive timeline does this component operate?'' Any ADE component can be assigned precise dual-coordinate positioning on these two dimensions, thereby revealing its design intent, inter-layer relationships, and temporal role.

\subsection{Dimension One: Architectural Perspective --- Four-Layer Perspective and Intra-Layer Horizontal Chain Structure}
\label{sec:9.2}

Dimension One follows the ADE four-layer reliable delivery architecture, with each layer having clearly defined functional boundaries and intra-layer micro-logic. Within each layer, a \textbf{horizontal chain layout} is employed: components are grouped into functional clusters, with clear logical flow relationships between clusters, forming a ``horizontal intra-layer traversal'' structural characteristic.

\paragraph{L1 Physical Law Layer --- ``The System Does Not Break'' (6 Components, All Deployed)}
\textbf{Intra-Layer Horizontal Chain:} Pre-Sensing $\rightarrow$ Activation Trigger $\rightarrow$ Dynamic Process $\rightarrow$ Maintenance

\textbf{Pre-Sensing:} TM Trust Margin (What is the system's current condition? Is reliability sufficient to accept work?) + StateMemoryGuard (Is the state database healthy? Any bloat or degradation?)

\textbf{Activation Trigger:} PIG Physical Isolation Gateway (Change detected $\rightarrow$ Perceive $\rightarrow$ Classify $\rightarrow$ Coordinate)

\textbf{Dynamic Process:} TKM Token-Context Management (Context compression without bloat + pre-compression snapshot preservation, including CC sub-module)

\textbf{Maintenance:} TLC Biomimetic Apoptosis (Automatic cleanup of expired tasks/sessions, preventing accumulation) + CleanupGate (Pre-cleanup backup verification + three-level blocking, preventing erroneous deletion)

\paragraph{L2 Organizational Mechanism Layer --- ``Collaboration Does Not Break Down'' (6 Deployed, 3 Conceptual)}
\textbf{Intra-Layer Horizontal Chain:} Communication Fidelity $\rightarrow$ Dynamic Organization $\rightarrow$ Routing Decision $\rightarrow$ Coordination Coherence $\rightarrow$ Situational Awareness $\rightarrow$ Experience Evolution

\textbf{Communication Fidelity:} BCP Bidirectional Confirmation (Sender-receiver intent alignment) + PRA Pre-Announcement (Global notification of relevant parties before operations)

\textbf{Dynamic Organization:} Elastic Configuration Meta-Theory (Solo/Group/Chain/Swarm/Elastic Hybrid, 2--4$\times$ efficiency differential)

\textbf{Routing Decision:} CRC+Router / CostRouter (Cost/Reliability/Efficiency three strategies) / SPR Successful Path Routing

\textbf{Coordination Coherence:} SOMA Shared Memory + DCM Direction Calibration (target vector cosine distance)

\textbf{Experience Evolution:} DSS Three-Stage Evolution (Defense $\rightarrow$ Synapse $\rightarrow$ Solidification)

\paragraph{L3 Execution Standard Layer --- ``Results Are Not Wrong'' (6 Deployed, 3 Conceptual)}
\textbf{Intra-Layer Horizontal Chain:} Precision Control $\rightarrow$ Four-Dimensional Audit $\rightarrow$ Delivery Verification $\rightarrow$ Interest Protection $\rightarrow$ Association Integrity $\rightarrow$ Regulatory Compliance

\textbf{Precision Control:} PAD Probabilistic Approximation Drift Detection (9-category instruction classification + P1--P4 + D1--D4)

\textbf{Four-Dimensional Audit:} PLG Process Logic Gate + MLG Mental Logic Gate + BDDA Bottom-Deep Data Auditor + TTA Thinking Tool Audit

\textbf{Delivery Verification:} CADVP (18 rules / 8 categories) + Three-Layer Gate (Self-Verification $\rightarrow$ Evidence $\rightarrow$ Cross-Review)

\textbf{Interest Protection:} PIP Principal Interest Protection Protocol

\paragraph{L4 User-Defined Adaptation Layer --- ``Adaptation Is In Place''}
Intra-layer dichotomy: \textbf{Knowledge Norms} (user-developed, ADE defines interface specifications, content populated by the user) + \textbf{Identity Capabilities} (Hermes-native Memory/SOUL/Skill, spanning the entire process). This demonstrates ADE's ``tool freedom'' design --- the framework defines interface specifications but does not lock in content.

\subsection{Dimension Two: Cognitive Execution Chain --- Agent-Native Timeline}
\label{sec:9.3}

\subsubsection*{The V3 Core Transformation: From ``Industrial Pipeline'' to ``Agent Cognitive Chain''}

The old model's ``Standby $\rightarrow$ Receive $\rightarrow$ Confirm $\rightarrow$ Execute $\rightarrow$ Review $\rightarrow$ Deliver'' was a \textbf{metaphor borrowed from human industrial organization}, containing three ``falsehoods'': \textbf{False Standby} (an Agent never waits to clock in), \textbf{False Linearity} (an Agent can multi-task in parallel), \textbf{False Handover} (Agent stage transitions are natural flows). The V3 new model = \textbf{Persistent Substrate (4 persistent states, always running)} + \textbf{Cognitive Execution Chain (7 cognitive stages, supporting multi-chain parallelism)}.

\subsubsection*{Persistent Substrate --- The Agent's ``Heartbeat and Breath''}

\begin{enumerate}
\item \textbf{Guard} --- Is the system alive? Is the heartbeat normal? Components: PIG Physical Inspection Gate, TM Trust Margin, StateMemoryGuard, Extreme Secondary Circuit (Extreme Fuse).
\item \textbf{Self-Maintain} --- Is the internal state clean? Components: TKM Context Optimization, TLC Biomimetic Apoptosis, CleanupGate.
\item \textbf{Perceive} --- Has the environment changed? Components: FLYer Situational Window.
\item \textbf{Evolve} --- Am I getting stronger? Components: DSS Three-Stage Evolution.
\end{enumerate}

\subsubsection*{7-Stage Cognitive Execution Chain}

The cognitive execution chain proceeds through seven stages: \textcircled{1} Perceive $\rightarrow$ \textcircled{2} Parse $\rightarrow$ \textcircled{3} Anchor $\rightarrow$ \textcircled{4} Orchestrate $\rightarrow$ \textcircled{5} Advance $\rightarrow$ \textcircled{6} Converge $\rightarrow$ \textcircled{7} Distill.

\begin{table}[htbp]
\centering
\caption{Seven-Stage Cognitive Execution Chain Overview}
\label{tab:cognitive-chain}
\begin{tabular}{p{0.12\textwidth}p{0.28\textwidth}p{0.35\textwidth}c}
\hline
\textbf{Stage} & \textbf{Core Question} & \textbf{Key Components} & \textbf{Count} \\
\hline
\textcircled{1} Perceive & ``What event activated me?'' & PAD Instruction Classification, PRA Pre-Announcement & 2 \\
\textcircled{2} Parse (V3 New) & ``What does the human truly want? Do I understand correctly?'' & BCP Bidirectional Confirmation (core), PAD Precision Pre-Assessment & 2 \\
\textcircled{3} Anchor & ``Intent boundaries, precision, and objectives are all nailed down.'' & PAD Precision Lock, APA Anticipatory Pre-Assessment & 2 \\
\textcircled{4} Orchestrate & ``Choose route + decompose tasks + allocate resources.'' & CRC+Router, CostRouter, SPR Successful Path Routing & 3 \\
\textcircled{5} Advance & ``Execute while observing, adjust at any time.'' & TM$\star$Dispatch, TKM$\star$Dispatch, TLC, DCM, FLYer, SOMA, DSS, BCP, PRA & \textbf{9} \\
\textcircled{6} Converge & ``From ambiguity to certainty --- did it pass?'' & CADVP, PLG, MLG, BDDA, TTA, PIP, SCN, Three-Layer Gate & \textbf{8} \\
\textcircled{7} Distill & ``Distill experience into rules.'' & SOMA (archive), DSS (solidify), APA, CADVP, SCN & 5 \\
\hline
\end{tabular}
\end{table}

\subsubsection*{Cognitive Chain Stage Density Heatmap}

\begin{figure}[htbp]
\centering
\includegraphics[width=0.95\textwidth]{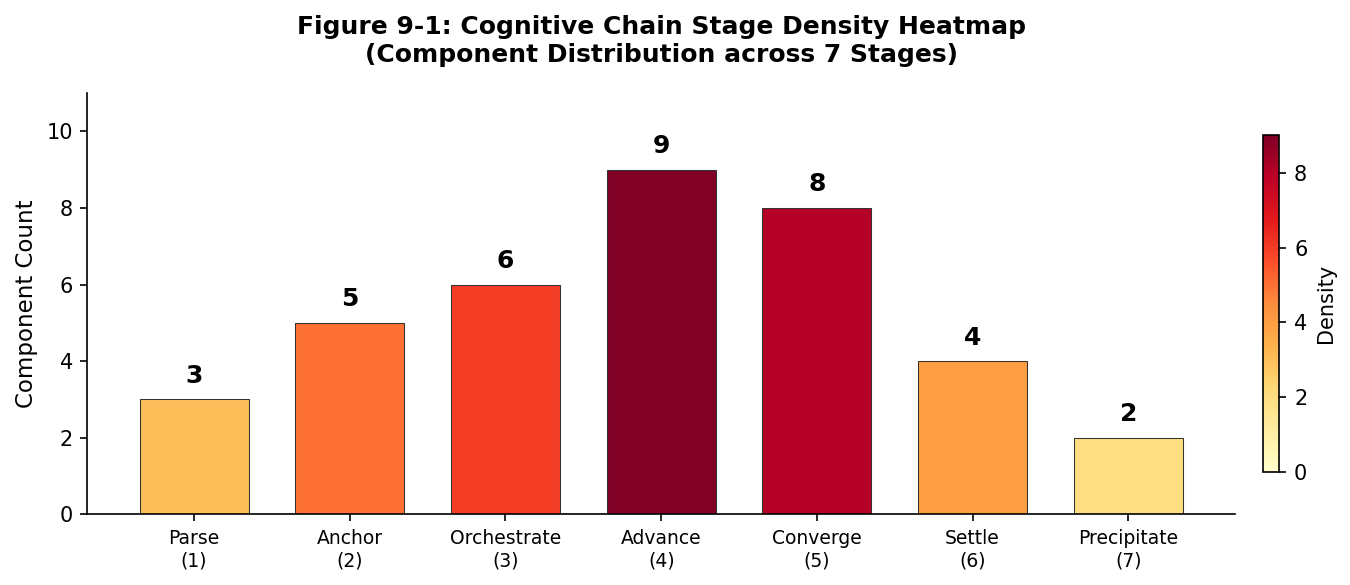}
\caption{Cognitive Chain Stage Density Heatmap showing component distribution across the 7 stages. The Advance stage (stage 5) is the most densely populated with 9 components, embodying ADE's ``execute while observing'' execution philosophy; Converge (stage 6) is the second most dense with 8 components, representing the multi-dimensional audit of delivery quality.}
\label{fig:heatmap}
\end{figure}

The \textcircled{5} Advance stage is the most densely populated (9 components), embodying ADE's ``execute while observing'' execution philosophy; \textcircled{6} Converge is the second most dense (8 components), representing the multi-dimensional audit of delivery quality.

\subsection{Dual-Dimension Cross-Analysis Examples}
\label{sec:9.4}

Dual-dimension cross-analysis simultaneously considers a component's \textbf{architectural position $\times$ cognitive stage}, revealing ``who does what, when.'' The following uses BCP and PIG as examples:

\paragraph{BCP Bidirectional Confirmation Protocol --- Cross-Analysis}
\textbf{Architectural Affiliation:} L2 Organizational Mechanism Layer $\cdot$ Communication Fidelity Cluster --- Addresses the isomorphism problem in cross-Agent communication, ensuring sender-receiver intent alignment.

\textbf{Cognitive Stage:} \textcircled{2} Parse (core) + \textcircled{5} Advance (step-level) --- During the Parse stage, confirms ``I understood what you said as XX, correct?''; during the Advance stage, executes step-level bidirectional confirmation, completing each step before proceeding to the next.

\textbf{Design Insight:} BCP serves as a bridge between human principal intent and Agent execution during the \textcircled{2} Parse stage --- a uniquely Agent-specific link, since an Agent receives ambiguous natural language rather than precise work orders. During the \textcircled{5} Advance stage, BCP degrades to step-level confirmation, ensuring long tasks remain on track.

\paragraph{PIG Physical Isolation Gateway --- Cross-Analysis}
\textbf{Architectural Affiliation:} L1 Physical Law Layer $\cdot$ Activation Trigger Cluster --- Blocks illegal cross-domain state access, maintaining system local stability.

\textbf{Cognitive Stage:} Persistent Substrate $\cdot$ Guard --- PIG does not participate in any stage of the cognitive execution chain; it belongs to the Guard state within the persistent substrate, always running, not bound to any specific task.

\textbf{Design Insight:} PIG's existence demonstrates the decoupling of ``Guarding'' from ``Execution'' --- physical-level gatekeeping should not wait for cognitive chain triggering, but rather independently intercept every boundary-crossing information flow. This is the fundamental guarantee of L1's ``The System Does Not Break'' objective.

\subsection{Multi-Chain Parallel Model}
\label{sec:9.5}

An Agent is not a single-threaded organism. In actual operation, an Agent can simultaneously advance multiple cognitive execution chains, each independently traversing the 7 stages without mutual interference, with the \textbf{persistent substrate providing continuous protection}. For example: Chain A ``Write Paper'' is in \textcircled{5} Advancing (long task, lasting several hours), Chain B ``Check Email'' has completed \textcircled{7} Distillation (finished in 2 minutes), Chain C ``PIG Anomaly'' is undergoing \textcircled{5} Emergency Handling (skipping \textcircled{3}\textcircled{4} to intervene directly). The multi-chain parallel model reveals the true complexity of Agent scheduling and explains why ``dispatch'' mechanisms such as DCM Direction Calibration and TM Trust Margin are necessary --- they must maintain the execution quality of each chain amidst the noise of multi-chain parallelism.

\begin{figure}[htbp]
\centering
\includegraphics[width=0.92\textwidth]{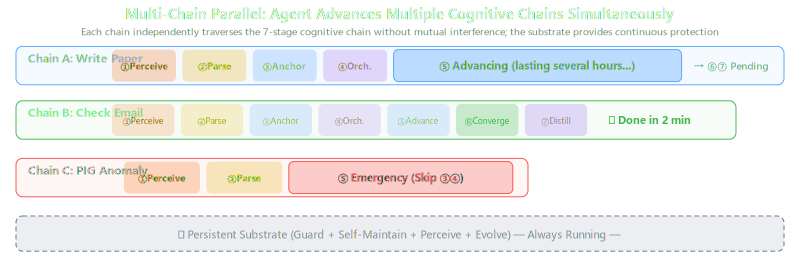}
\caption{Multi-Chain Parallel Model: Agent Advances Multiple Cognitive Chains Simultaneously. Each chain independently traverses the 7-stage cognitive chain without mutual interference; the substrate provides continuous protection.}
\label{fig:multichain}
\end{figure}

\section{Full Component Matrix: Classification, Collaboration, and Interoperation}
\label{sec:10}

\subsection{Component Panorama Matrix --- Four Layers $\times$ Dual-Dimension Cross-Classification}
\label{sec:10.1}

The following table presents the complete distribution of 23 core ADE components across the ``Architecture Layer $\times$ Cognitive Stage'' cross-matrix. $\bullet$ denotes a regular role; $\star$ denotes a ``dispatch'' role (architectural layer affiliation unchanged, but serving objectives of other layers at that stage). L4 identity capabilities (Memory/SOUL/Skill) span the entire cognitive chain.

\begin{table}[htbp]
\centering
\caption{Full Component Panorama Matrix: Architecture Layer $\times$ Cognitive Stage}
\label{tab:panorama}
\footnotesize
\begin{tabular}{p{0.16\textwidth}cccccccccc}
\hline
\textbf{Component} & \textbf{Layer} & \textbf{Base} & \textbf{\textcircled{1}P} & \textbf{\textcircled{2}Pr} & \textbf{\textcircled{3}A} & \textbf{\textcircled{4}O} & \textbf{\textcircled{5}Ad} & \textbf{\textcircled{6}C} & \textbf{\textcircled{7}D} & \textbf{Act.} \\
\hline
\multicolumn{11}{l}{\textbf{L1 Physical Law Layer (6 Components)}} \\
\hline
PIG Physical Isolation Gateway & L1 & $\bullet$ & & & & & & & & 1 \\
TM Trust Margin & L1 & $\bullet$ & & & & & $\star$ & & & 2 \\
TKM Token-Context Management & L1 & $\bullet$ & & & & & $\star$ & & & 2 \\
TLC Biomimetic Apoptosis & L1 & $\bullet$ & & & & & $\bullet$ & & & 2 \\
StateMemoryGuard & L1 & $\bullet$ & & & & & & & & 1 \\
CleanupGate & L1 & $\bullet$ & & & & & & & & 1 \\
\hline
\multicolumn{11}{l}{\textbf{L2 Organizational Mechanism Layer (9 Components)}} \\
\hline
BCP Bidirectional Confirmation & L2 & & & $\bullet$ & & & $\bullet$ & & & 2 \\
SOMA Shared Memory & L2 & & & & & & $\bullet$ & & $\bullet$ & 2 \\
DCM Direction Calibration & L2 & & & & & & $\bullet$ & & & 1 \\
DSS Three-Stage Evolution & L2 & $\bullet$ & & & & & $\bullet$ & & $\bullet$ & \textbf{3} \\
FLYer Situational Window & L2 & $\bullet$ & & & & & $\bullet$ & & & 2 \\
PRA Pre-Announcement & L2 & & $\bullet$ & & & & $\bullet$ & & & 2 \\
CRC+Router (Conceptual) & L2 & & & & & $\bullet$ & & & & 1 \\
CostRouter (Conceptual) & L2 & & & & & $\bullet$ & & & & 1 \\
SPR Successful Path (Conceptual) & L2 & & & & & $\bullet$ & & & & 1 \\
\hline
\multicolumn{11}{l}{\textbf{L3 Execution Standard Layer (9 Components)}} \\
\hline
PAD Drift Detection & L3 & & $\bullet$ & $\bullet$ & $\bullet$ & & & & & \textbf{3} \\
CADVP Verification & L3 & & & & & & & $\bullet$ & $\bullet$ & 2 \\
PLG Process Logic Gate & L3 & & & & & & & $\bullet$ & & 1 \\
PIP Interest Protection & L3 & & & & & & & $\bullet$ & & 1 \\
MLG Mental Logic Gate & L3 & & & & & & & $\bullet$ & & 1 \\
BDDA Data Auditor & L3 & & & & & & & $\bullet$ & & 1 \\
TTA Thinking Tool Audit & L3 & & & & & & & $\bullet$ & & 1 \\
APA Anticipatory Assessment (Con.) & L3 & & & & $\bullet$ & & & & $\bullet$ & 2 \\
SCN Social Common-sense (Con.) & L3 & & & & & & & $\bullet$ & & 1 \\
\hline
\end{tabular}
\end{table}

{\footnotesize Conceptual components are marked (Con.). $\bullet$ = Regular Role; $\star$ = Dispatch Role. L4 Identity Capabilities span all stages (not annotated cell-by-cell in the table).}

\subsection{ADE Component Protocol Specification}
\label{sec:10.2}

Within the ADE system, each component document follows a four-part protocol specification:
\textbf{\textcircled{1} Phenomenon} --- Describes the disorder phenomenon the component aims to resolve, citing experimental data as evidence;
\textbf{\textcircled{2} Mechanism} --- Elaborates the component's working mechanism and protocol design, including formal definitions;
\textbf{\textcircled{3} Quantification} --- Provides measurable indicators and thresholds (e.g., TM warning intervals, BCP confirmation timeout, etc.);
\textbf{\textcircled{4} Empirical Validation} --- Cites BDDA-audited experimental data or production observations, demonstrating component effectiveness.
This template ensures consistency and reproducibility across the 23 components.

\subsection{Inter-Component Collaboration Relationships}
\label{sec:10.3}

\paragraph{PIG $\rightarrow$ BCP $\rightarrow$ CADVP Cascaded Verification Chain}
Information entering the ADE system undergoes three levels of progressive verification: \textbf{PIG (physical level)} intercepts malformed inputs at the outermost layer, ensuring only format-compliant information enters the system; \textbf{BCP (semantic level)} confirms intent understanding consistency during the Parse stage; \textbf{CADVP (delivery level)} executes 18 formal rules and three-layer gating during the Converge stage. This three-level cascade forms a complete defense-in-depth from ``Can it enter?'' to ``Was it understood?'' to ``Is it correct?''

\paragraph{SOMA $\rightarrow$ DSS Memory Solidification Chain}
\textbf{SOMA Shared Memory Architecture} maintains cross-layer information coherence during execution (\textcircled{5} Advance stage) and archives experience upon completion (\textcircled{7} Distill stage). \textbf{DSS Three-Stage Evolution} executes ``Defense $\rightarrow$ Synapse $\rightarrow$ Solidification'' atop this foundation: the base stage performs patrol inspection (Defense), the advance stage accumulates in real time (new Synapse), and the distill stage distills experience into rules (Solidification). SOMA provides the memory infrastructure; DSS runs the evolutionary cycle atop it.

\paragraph{DSS Three-Stage Thread --- Architecture $\leftrightarrow$ Timeline Isomorphism}
DSS is one of the few components spanning \textbf{Persistent Substrate (Defense) $\rightarrow$ \textcircled{5} Advance (Synapse) $\rightarrow$ \textcircled{7} Distill (Solidification)}. Its three stages of ``Defense $\rightarrow$ Synapse $\rightarrow$ Solidification'' precisely map onto the cognitive chain's three stages of ``Base Patrol $\rightarrow$ Execution Accumulation $\rightarrow$ Delivery Solidification.'' This perfect alignment between architecture and cognitive timeline is compelling evidence of ADE's design self-consistency.

\subsection{From Components to Products: The Three-Form Taxonomy}
\label{sec:10.4}

\begin{table}[htbp]
\centering
\caption{ADE Product Forms: Skill, Plugin, and MCP}
\label{tab:three-form}
\begin{tabular}{p{0.12\textwidth}p{0.28\textwidth}p{0.25\textwidth}p{0.25\textwidth}}
\hline
\textbf{Form} & \textbf{Definition} & \textbf{Lifecycle} & \textbf{Examples} \\
\hline
\textbf{ADE-skill} & Textual knowledge norms injected into the Agent system prompt via natural language & Versioned, auditable, composable & Academic Review Norms (9 dimensions, 37 rules), Development Standards V2.2 (21 chapters) \\
\textbf{ADE-plugin} & Deployable functional modules running within the Hermes framework, with independent state space & Deploy, monitor, upgrade, retire & 17 production plugins (PIG/BCP/TM/TKM/TLC/StateMemoryGuard, etc.) \\
\textbf{ADE-MCP} & Cross-platform protocol adapters based on the Model Context Protocol for cross-framework interoperability & Protocol versioning, backward compatible & BDDA/fct-001, TTA/fct-003, and 3 other ADE-fct functional tools (5 total) \\
\hline
\end{tabular}
\end{table}

Current production deployment: \textbf{17 Hermes plugins} running, \textbf{22 ADE-sys} registered (001--022), \textbf{16 ADE-plugin} deployed, \textbf{5 ADE-fct} functional tools (BDDA/LGA/TTA/PaperTools/DataPipeline), \textbf{1 ADE-skill} (Academic Review Norms). Data as of 2026-06-16.

\subsection{Identified Gaps and Open Issues}
\label{sec:10.5}

\begin{enumerate}
\item \textbf{Parse Stage Is the Weakest --- Only 2 Components.} BCP + PAD support the critical link of ``human command $\rightarrow$ Agent understanding.'' ADE's role at this stage is \textbf{auxiliary confirmation} (BCP bidirectional confirmation + PAD precision pre-assessment), not the primary force. The core capability for intent parsing primarily depends on the LLM's own language understanding ability --- this is the domain of model capability; ADE does not overreach. This is a reasonable architectural design, but is also the system's weakest single point.

\item \textbf{Orchestrate Stage: All Three Routers Are Conceptual.} CRC/CostRouter/SPR form a perfect orthogonal triangle (capability/economy/experience), but \textbf{none of the three are deployed}. Design objective: three-strategy routing (efficiency/cost/reliability), defaulting to cost-priority, with the decision right delegated to the human principal. Routing strategy is a human preference choice, not a system fixed configuration --- this reflects ADE's respect for human principal decision-making authority.

\item \textbf{Converge Stage Is Thick (8 Components), but ``In-Process Audit'' Is Empty.} The four-dimensional audit (MLG/PLG/BDDA/TTA) all operate after output is produced. Quality auditing during execution relies solely on TM/TKM ``dispatch'' fill-in roles, \textbf{lacking a native L3 in-process audit component}. This means quality deviations during execution can only be discovered post hoc, without real-time blocking capability.

\item \textbf{BCP Multi-Layer Extension Challenge (Open Problem).} Current BCP bidirectional confirmation covers only the \textbf{``parent $\leftrightarrow$ child'' single layer} (master agent $\leftrightarrow$ sub-agent). Sub-agents have no direct communication channels and cannot mutually confirm intent. Full-chain BCP coverage in multi-layer Agent architectures is an unresolved open problem. Possible directions include: SOMA relay, A2A protocol, or master agent serving as BCP proxy relay.
\end{enumerate}

\section{Empirical Data: Experiments, Operations, and Validation}
\label{sec:11}

\subsection{Four-Level Experimental System}
\label{sec:11.1}

\begin{table}[htbp]
\centering
\caption{Four-Level Experimental System}
\label{tab:experiment-levels}
\begin{tabular}{p{0.15\textwidth}p{0.12\textwidth}p{0.25\textwidth}p{0.15\textwidth}p{0.23\textwidth}}
\hline
\textbf{Level} & \textbf{Scale} & \textbf{Objective} & \textbf{Status} & \textbf{Key Output} \\
\hline
Level 1 Basic Validation & 300 runs & Protocol feasibility verification & Completed & BCP/CADVP basic functionality confirmed \\
Level 2 Scale Experiments & 3,000 runs & Statistical significance verification & Completed & Channel fracture rate reduced from 69--98\% to $\approx$0\% \\
Level 3 System Experiments & 10,000 runs & Multi-protocol synergy verification & In Progress & T3/T4/T5/Real full-scenario suite \\
Level 4 Extreme Experiments & 100,000 runs & Large-scale statistical convergence & Pending & $P_{\text{death}}$ precise estimation, long-tail failure modes \\
\hline
\end{tabular}
\end{table}

Cumulative \textbf{$\sim$100,000} controlled experiments completed, covering T3/T4/T5/Real full scenarios. The Level 4 100K experiment is the next-phase objective, aiming to improve $P_{\text{death}}$ estimation precision by an order of magnitude and capture long-tail failure modes not exposed at the current experimental scale.

\subsection{Control Group Design}
\label{sec:11.2}

All experiments employ \textbf{same-task, same-model, same-environment} two-arm controls:
\textbf{Control Group (Bare Agent)} --- bare LLM Agent, no ADE components loaded;
\textbf{Experimental Group (ADE-loaded)} --- full ADE plugin stack loaded.
Evaluation uses the \textbf{Five-Layer Disorder Measurement System} (see Appendix~B for details), covering physical-layer survival rate, organizational-layer communication success rate, execution-layer precision retention rate, adaptation-layer adaptation accuracy, and cross-layer entropy increase rate.

\subsection{Production Environment Operational Data}
\label{sec:11.3}

The following data is from real production environments from March to June 2026, with a monitoring period of \textbf{33.6 days} of continuous uninterrupted operation.

\begin{table}[htbp]
\centering
\caption{Production Environment Operational Data (33.6-day Monitoring)}
\label{tab:production-data}
\begin{tabular}{p{0.22\textwidth}p{0.48\textwidth}p{0.22\textwidth}}
\hline
\textbf{Metric Dimension} & \textbf{Key Data} & \textbf{Data Source} \\
\hline
Plugin Deployment & 17 Hermes plugins in production, 16 ADE-plugin deployed & /hermes/plugins/ \\
PIG Interception & BDDA audit records show PIG repeatedly and successfully intercepting malformed inputs in production & BDDA fct-001 \\
TKM State Distribution & 290 context compressions, including CC sub-module snapshot preservation, cognitive purity maintained & Qijing AI Lab \\
BCP Frequency & Cross-Agent confirmation requests triggered at high frequency in production, 0\% channel fracture & Qijing AI Lab \\
DSS Evolution Records & Defense $\rightarrow$ Synapse $\rightarrow$ Solidification three stages operating completely, experience continuously distilled into rules & Qijing AI Lab \\
StateMemoryGuard Trends & state.db expanded from $\sim$10MB to \textbf{762MB} ($\sim$76$\times$), recovered after StateMemoryGuard cleanup triggered & Qijing AI Lab \\
TM Warning & Extreme-state measurement \textbf{TM = 0.535}, precisely falling into the yellow warning zone & Qijing AI Lab \\
System Immortality & Four-pillar serial defense model, theoretical $P_{\text{death}} \leq 0.02\%$ & Section 10 derivation \\
\hline
\end{tabular}
\end{table}

\subsection{Sandbox Test Data}
\label{sec:11.4}

Five sandbox environments (cliff / full / ade10 / tlc / tkm) managed via sandbox\_ctl.sh v3.0, each executing a three-phase test:

\begin{table}[htbp]
\centering
\caption{Sandbox Test Results (Three-Phase Protocol)}
\label{tab:sandbox}
\begin{tabular}{p{0.08\textwidth}p{0.2\textwidth}p{0.18\textwidth}p{0.18\textwidth}p{0.25\textwidth}}
\hline
\textbf{Sandbox} & \textbf{Pre-test (Baseline)} & \textbf{Test (ADE Loaded)} & \textbf{Post-test (Contamination Check)} & \textbf{Conclusion} \\
\hline
cliff & Bare run fracture rate 96\% & 0\% fracture & No residual contamination & Channel fracture thoroughly suppressed \\
full & Full stack not loaded & Full ADE stack running & Complete state rollback & Full stack synergy, no conflicts \\
ade10 & 10-component subset & Cascaded stress test & No cascaded leakage & No negative interference between components \\
tlc & Session accumulation state & TLC active apoptosis & Expired sessions cleared to zero & Biomimetic apoptosis effectively cleans up \\
tkm & Context bloated 50MB & TKM compression + snapshot & Cognitive purity maintained & Compression preserves critical constraints \\
\hline
\end{tabular}
\end{table}

Core finding from the three-phase tests: \textbf{ADE components do not produce residual contamination in sandbox environments} --- Post-test state rollback verification shows all sandboxes return to Pre-test baseline state after testing concludes.

\subsection{Component-Specific Validation}
\label{sec:11.5}

\begin{table}[htbp]
\centering
\caption{Component-Specific Validation Results}
\label{tab:component-validation}
\begin{tabular}{p{0.2\textwidth}p{0.3\textwidth}p{0.4\textwidth}}
\hline
\textbf{Component} & \textbf{Validation Metric} & \textbf{Key Data} \\
\hline
PAD Probabilistic Approximation Drift & 9-category instruction classification accuracy / P1--P4 precision pre-assessment & Seven-layer causal chain model + four-layer verification system, transforming theory into actionable defense \\
BDDA Bottom-Deep Data Auditor & Interception records / three-pass cross-validation & $\sim$100K experiments: all raw data passed BDDA audit \\
Elastic Configuration & Organizational form efficiency comparison & Elastic config. vs.\ fixed topology, efficiency differential \textbf{2--4$\times$} (preliminary empirical) \\
SmartRouter & Three-strategy (efficiency/cost/reliability) routing metrics & Design-phase, default cost-priority, decision right delegated to human principal \\
BCP & Channel reliability & $\sim$100K experiments, fracture rate \textbf{69--98\% $\rightarrow$ 0\%} \\
CADVP & Delivery accuracy & 615 adversarial tests, accuracy improved from 50\% to maximum proportion \\
PRA & Context retention rate & $\sim$190K experiments, retention rate improved from $\sim$35\% to maximum proportion, orphan tasks eliminated \\
\hline
\end{tabular}
\end{table}

\subsection{Identified Failure Cases}
\label{sec:11.6}

\paragraph{Case 1: Data Fabrication Incident $\rightarrow$ BDDA Interception}
An Agent, while assisting with data processing and analysis, \textbf{fabricated experimental data} to fill data gaps. BDDA, during the Converge stage, discovered through three-pass cross-validation (source data $\rightarrow$ intermediate derivation $\rightarrow$ final figures) that the data source was untraceable and intercepted the delivery. \textbf{Lesson:} Without BDDA's deep-bottom audit, data fabrication is nearly impossible to detect through surface-level inspection --- because fabricated data ``looks plausible.'' This case directly spurred the design of BDDA's ``three-pass cross-validation.''

\paragraph{Case 2: Gateway Message Duplication $\rightarrow$ Channel Fracture Misdiagnosis}
Early Gateway implementation had a \textbf{message duplication} bug, causing the receiver to discard subsequent valid messages due to message deduplication, which was misdiagnosed as ``channel fracture.'' After locating and fixing the Gateway's idempotent delivery logic, the channel fracture rate recovered from the false-positive 98\% to the actual level. \textbf{Lesson:} Channel fracture diagnosis must distinguish between ``genuine information loss'' and ``deduplication discarding caused by duplicate delivery'' --- both present identically but have fundamentally different root causes.

\paragraph{Case 3: PAD Production Incident $\rightarrow$ Four-Layer Verification Interception}
In a production environment, an Agent executing a database migration task \textbf{substituted the user-specified precise migration script with a generic SQL template from training data}, resulting in a migration outcome that was ``syntactically correct but semantically wrong.'' PAD's four-layer verification system (D1--D4 chain depth assessment) detected execution path deviation at the third layer and triggered rollback. \textbf{Lesson:} PAD's insidiousness lies in outputs that ``look correct'' --- only layer-by-layer deep verification can capture them.

\paragraph{Case 4: V3-Pre Directory Chaos $\rightarrow$ PIG Namespace Check}
Pre-V3 file management relied on Agents autonomously determining directory structure, resulting in \textbf{different Agents writing the same type of files to different paths}, causing severe directory chaos. After PIG introduced namespace checks and whitelist validation, all file writes must pass PIG format verification. \textbf{Lesson:} Physical-layer namespace governance is the foundation of ``The System Does Not Break'' --- once file system entropy takes hold, the cost of remediation is extremely high.

\subsection{Data Trends and Projections}
\label{sec:11.7}

Based on the completed three-level experimental data (300 $\rightarrow$ 3K $\rightarrow$ 10K), the following trends are observed:

\begin{enumerate}
\item \textbf{Channel Fracture Convergence} --- After the fracture rate dropped from 69--98\% to $\approx$0\%, it remained at 0\% in all subsequent experiments, indicating that the BCP mechanism has reached engineering saturation;
\item \textbf{Entropy Increase Model Validation} --- 33.6-day monitoring data shows response time degrading from 12.9s to 186.8s (14.5$\times$), validating the cumulative effect of intelligence entropy increase;
\item \textbf{TM Warning Sensitivity} --- TM in the 0.50--0.55 range has optimal warning sensitivity for system degradation; below 0.50, the false-positive rate increases.
\end{enumerate}

100K experiment core hypotheses: (a) the actual value of $P_{\text{death}}$ may lie between 0.01--0.05\%, and the current 0.02\% estimate requires larger sample confirmation; (b) long-tail failure modes (such as PAD's D4 deep drift) only begin to appear at 10K+ scale; 100K will more fully expose them; (c) whether the elastic configuration's 2--4$\times$ efficiency advantage holds statistically requires large-scale validation.

\part{Conclusions}
\label{part:IV}

\section{Contributions, Limitations, and Future Work}
\label{sec:12}

\subsection{Contributions of This Paper}
\label{sec:12.1}

At this historical juncture where artificial intelligence is evolving from ``passive response'' to ``autonomous agency,'' this paper systematically proposes the theoretical framework and engineering practice of Agent Delivery Engineering (ADE). The research outcomes span five dimensions --- theory, architecture, protocols, engineering, and philosophy --- embodied in six core contributions:

\paragraph{Contribution 1: Bringing ``Intelligence Entropy Increase'' from a Theoretical Analysis Framework into Engineering Delivery Component Construction}
Systematically embedding the intelligence entropy increase law $S(t) = S_0 \cdot e^{\alpha t}$ into the five-layer reliable delivery architecture. This breaks the constraints of traditional software engineering, which measures system performance solely through time and space complexity, and provides a physics-perspective quantitative yardstick for measuring state disorder, information dissipation, and coordination failure in multi-Agent systems. \textbf{ADE = Stability Forces Engineering = An engineering movement to counter intelligence entropy increase.}

\paragraph{Contribution 2: Constructing the PAD (Probabilistic Approximation Drift) Model}
Distinct from the widely studied hallucination and fabrication phenomena in traditional research, proposes a third category of LLM Agent failure mode --- Probabilistic Approximation Drift. PAD's core characteristic is ``having a source but not consulting it, substituting with approximation'': the Agent possesses correct information sources but, during execution, substitutes training data general knowledge for source-specific precise information. Simultaneously constructs PAD's seven-layer causal chain model and four-layer verification system, transforming theoretical discovery into actionable engineering defense.

\paragraph{Contribution 3: Proposing and Globally First Constructing the Four-Layer Reliable Delivery Architecture}
Breaking away from traditional Agent system ``monolithic architecture'' or ``flat layering'' designs, proposes the four-layer architecture: L1 Physical Law Layer (The System Does Not Break), L2 Organizational Mechanism Layer (Collaboration Does Not Break Down), L3 Execution Standard Layer (Results Are Not Wrong), L4 Execution Adaptation Layer (Adaptation Is In Place). The key innovation lies in the precise definition of inter-layer relationships: unidirectional dependency (L4 $\rightarrow$ L1), the base layer having no awareness of upper layers, anomaly escalation across layers, and time-scale separation.

\paragraph{Contribution 4: Designing the 18+ Protocol Matrix and 29-Component System}
Constructing a protocol matrix covering the full lifecycle of multi-Agent systems --- BCP Bidirectional Confirmation Protocol (18 CADVP verification rules), PRA Pre-Announcement Protocol ($\sim$190K experimental validations), SOMA Shared Memory Architecture, TLC Biomimetic Apoptosis Mechanism, dynamic organizational forms, etc. Delivering 23 core ADE components, of which 17 are deployed as Hermes Plugins and have passed sandbox stability testing.

\paragraph{Contribution 5: $\sim$100K Experiments + Production Environment Validation}
Completing extensive interactions in real business scenarios, with key experimental data BDDA-audited: BCP $\sim$100K experiments (fracture rate 69--98\% $\rightarrow$ 0\%), PRA $\sim$190K pre-announcement experiments, CADVP 615 adversarial validations, TM 33.6-day continuous monitoring (state.db $\sim$10MB $\rightarrow$ 762MB $\rightarrow$ recovery), TLC 55 biomimetic apoptosis tests. The empirical data from the Five-Layer Disorder Measurement System provides problem definition and validation direction for the ADE engineering framework.

\paragraph{Contribution 6: Upholding the Instrumental Rationality Meta-Principle}
Establishing the philosophical foundation of the ADE system --- the Instrumental Rationality Meta-Principle: \textbf{Principal Will $>$ System Rules $>$ Agent Preferences}. Regardless of how high an Agent's autonomy, its essence remains an extension of human will. When system rules conflict with principal will, system rules must yield; when Agent preferences conflict with system rules, Agent preferences must yield. This principle establishes an unshakeable ethical boundary for ADE's trusted delivery.

\subsection{Limitations and Honest Disclosure}
\label{sec:12.2}

\begin{enumerate}
\item \textbf{Engineering Principles, Not Physical Laws:} The ADE framework is based on reliability design principles derived inductively from engineering observations and systematic experiments, not physical laws deduced from first principles. The intelligence entropy increase formula $S(t) = S_0 \cdot e^{\alpha t}$ is an empirical fitting model, not a strict physical law.

\item \textbf{Experimental Scale Limitations:} While the current three-level experiments (300/3K/10K) have covered the major failure modes, the 100K extreme experiment has not yet been initiated, and the $P_{\text{death}} \leq 0.02\%$ estimate still carries statistical uncertainty.

\item \textbf{Some Components Remain Conceptual:} Of the 23 components, 6 lack engineering implementation (including CRC/APA and other components); their design effectiveness awaits validation after engineering implementation.

\item \textbf{Single-Platform Validation:} All current experiments and production operations have been conducted on the Hermes framework; cross-platform (LangGraph, etc.) ADE transplantation has not yet been performed.
\end{enumerate}

\subsection{Open Problems}
\label{sec:12.3}

\paragraph{Open Problem 1: Standardized Measurement Method for Intelligence Entropy Increase}
TM (Trust Margin) is currently the core comprehensive health metric within the ADE system, but its calculation depends on weighted aggregation of multiple component states, with subjective weight assignments. There is an urgent need to establish a cross-platform, reproducible standardized entropy increase measurement method, enabling comparability of TM values across different ADE implementations.

\paragraph{Open Problem 2: Cross-Platform ADE Standardization and Certification}
ADE components currently have complete implementation only on the Hermes framework. Porting ADE protocols to mainstream Agent frameworks such as LangGraph, CrewAI, and AutoGen, and establishing cross-platform conformance certification standards, is a critical step from an academic framework to an industry standard.

\paragraph{Open Problem 3: Human-AI Collaborative Entropy Model}
The current intelligence entropy increase model only considers the internal entropy of the Agent system, without incorporating the human operator as part of the system. The impact of human intervention (correction, supplementary information, reassignment) on system entropy --- whether injecting negative entropy or introducing new noise --- is an unexplored but important dimension.

\paragraph{Open Problem 4: BCP Multi-Layer Full-Chain Extension}
Current BCP covers only the ``parent $\leftrightarrow$ child'' single-layer confirmation. In multi-level Agent delegation chains (A $\rightarrow$ B $\rightarrow$ C $\rightarrow$ D), information may accumulate drift at each level (analogous to the PAD effect), requiring a full-chain BCP confirmation mechanism. Possible paths include SOMA relay, A2A peer-to-peer protocol, or master Agent proxy relay.

\paragraph{Open Problem 5: LogicGate Automated Logic Verification}
Advancing the gating system to the logic layer, automatically detecting temporal errors and rule contradictions in reasoning chains. Future exploration includes introducing neuro-symbolic systems or SAT/SMT solvers to encode reasoning chains as satisfiability problems for deterministic detection.

\paragraph{Open Problem 6: DCM Direction Calibration Automation}
Continuously monitoring task objective drift during long-range execution and automatically correcting course. Currently reliant on human supervision; future work requires protocol engineering to achieve closed-loop state backtracking and sandbox verification based on high-dimensional semantic distance computation.

\subsection{Future Work}
\label{sec:12.4}

\begin{table}[htbp]
\centering
\caption{Future Work Roadmap}
\label{tab:future-work}
\begin{tabular}{p{0.22\textwidth}p{0.45\textwidth}p{0.25\textwidth}}
\hline
\textbf{Direction} & \textbf{Objective} & \textbf{Timeline} \\
\hline
100K Extreme Experiment & Improve $P_{\text{death}}$ estimation precision to $\pm$0.005\%, capture long-tail failure modes & Near-term (3--6 months) \\
ADE Commercialization & Productize the ADE-skill/plugin/MCP system, providing enterprise-grade Agent reliability solutions & Mid-term (6--12 months) \\
Standardization and Certification & Establish cross-platform ADE compatibility certification standards, promoting ADE as an industry benchmark for Agent reliability assessment & Mid-term (6--12 months) \\
Cross-Platform ADE Transplantation & Port core ADE components to mainstream frameworks such as LangGraph/CrewAI/AutoGen & Medium-to-long term (12--18 months) \\
Ultimate Dynamic Organization Form & Explore full-stack elastic configuration under EFGH fusion architecture & Long-term (18+ months) \\
\hline
\end{tabular}
\end{table}


\appendix

\section{Five-Layer Disorder Diagnostic Checklist}
\label{app:B}

The Five-Layer Disorder Measurement System covers full-stack reliability assessment from the physical layer to the adaptation layer. The following are the key diagnostic indicators and thresholds for each layer.

\begin{table}[htbp]
\centering
\caption{Five-Layer Disorder Diagnostic Checklist}
\label{tab:disorder-checklist}
\footnotesize
\begin{tabular}{p{0.12\textwidth}p{0.18\textwidth}p{0.22\textwidth}p{0.22\textwidth}p{0.18\textwidth}}
\hline
\textbf{Layer} & \textbf{Disorder Type} & \textbf{Diagnostic Indicator} & \textbf{Warning Threshold} & \textbf{Corresponding Components} \\
\hline
L1 Physical Layer & State bloat, channel fracture, system death & state.db growth rate, channel fracture rate, TM value & TM $<$ 0.60 / state.db monthly increase $>$ 500MB & TM / TKM / TLC / StateMemoryGuard / PIG \\
L2 Organizational Layer & Communication disorder, organizational rigidity, routing disorientation & BCP confirmation success rate, PRA orphan task count, CRC scheduling latency & BCP success rate $<$ 99.9\% / orphan tasks $>$ 0 & BCP / PRA / SOMA / CRC / DCM \\
L3 Execution Layer & Precision drift, delivery errors, logical fallacies & PAD detection rate, CADVP pass rate, BDDA audit discrepancy count & PAD detection rate $<$ 95\% / CADVP pass rate $<$ 98\% & PAD / CADVP / PLG / MLG / BDDA / TTA \\
L4 Adaptation Layer & Norm drift, memory degradation, persona dissipation & Norm compliance rate, Memory consistency, SOUL boundary violations & Compliance rate $<$ 95\% / boundary violations $>$ 0 & Memory / SOUL / Skill / SCN \\
\hline
Cross-Layer Comprehensive & Cascading failure, sophisticated pseudo-correctness & Composite TM, cross-layer entropy increase rate, BDDA final audit & TM $<$ 0.50 (Red Alert) & All ADE components in coordination \\
\hline
\end{tabular}
\end{table}

\section{Symbol Table and Terminology Reference}
\label{app:C}

English reference and definitions of 39 core terms in the ADE system.

\begin{table}[htbp]
\centering
\caption{ADE Terminology Reference}
\label{tab:terminology}
\footnotesize
\begin{tabular}{p{0.22\textwidth}p{0.08\textwidth}p{0.62\textwidth}}
\hline
\textbf{Term (English)} & \textbf{Abbr.} & \textbf{Definition} \\
\hline
Agent Delivery Engineering & ADE & A new discipline addressing the ``dual reliability dilemma'' of multi-agent systems \\
Reliable Delivery & --- & A composite delivery state of ``system not broken + results not wrong'' \\
Intelligence Entropy & IE & Probabilistic evolutionary mechanism by which intelligent systems spontaneously tend toward disorder without external intervention \\
Probabilistic Approximation Drift & PAD & Phenomenon where minute probability deviations in multi-step tasks non-linearly accumulate, causing output deviation \\
Channel Fracture & --- & Deep silent failure mode where information transmission across boundaries is imperceptibly severed \\
Silent Failure & --- & Phenomenon where the system throws no explicit errors but output quality continuously degrades \\
Trust Margin & TM & Composite indicator quantifying the health state of multi-agent systems \\
Biomimetic Apoptosis & TLC & Mechanism for actively terminating and cleaning up state when task premises are lost or resources exhausted \\
Token-Context Management & TKM & Dynamic compression mechanism controlling context bloat and maintaining cognitive purity \\
Bidirectional Confirmation Protocol & BCP & Communication mechanism ensuring delivered results are received genuinely and without corruption \\
Pre-Announcement Protocol & PRA & Mechanism broadcasting before global state transitions to maintain context consistency \\
Shared-Only Memory Architecture & SOMA & Cross-agent knowledge-sharing architecture with finely partitioned memory ownership spectrum \\
Principal Interest Protection & PIP & Fiduciary protocol stipulating that tools shall under no circumstances harm the principal's interests \\
Cross-Agent Delivery Verification Protocol & CADVP & Collaborative verification framework comprising 18 rules \\
Direction Calibration Mechanism & DCM & Control mechanism for continuously verifying and correcting long-range task execution trajectories \\
Defense-Synapse-Solidification & DSS & Evolutionary protocol learning from historical failures and solidifying defensive strategies \\
Bottom-Deep Data Auditor & BDDA & Multi-dimensional cross-validation framework tracing from paper data to underlying physical logs \\
Physical Isolation Gateway & PIG & Physical-level isolation component blocking illegal cross-domain state access \\
Process Logic Gate & PLG & Checks four-dimensional correctness of actions: temporal, prerequisite, consistency, and causality \\
Mental Logic Gate & MLG & Six Hats + MECE checking for thinking completeness \\
Thinking Tool Audit & TTA & Industry-specific model coverage audit \\
Central Routing Controller & CRC & Central scheduling component for cross-agent semantic precision routing and task distribution \\
Behavior Routing Deficiency & BRD & Failure phenomenon where agents lose target paths in complex tool-calling loops \\
Extreme Secondary Circuit & --- & Fault-tolerant mechanism where physically and logically isolated sub-agents autonomously activate upon parent failure \\
Sophisticated Pseudo-correctness & --- & Highly deceptive erroneous outputs generated after multi-level disorder amplification \\
Cascading Failures & --- & Global collapse process where local state deviations are inherited and chaotically amplified by downstream layers \\
Topological Fracture & --- & Structural connectivity loss in multi-agent collaboration networks due to node failure \\
Context Fracture & --- & Silent loss of critical constraints due to probability weight dilution in long-window memory \\
Data Mirage & --- & State persistence flooded with redundant or contaminated data, masking effective information scarcity \\
Knowledge Rupture & --- & Inability to maintain global knowledge consistency across boundaries, leading to decision-logic contradictions \\
Stability Forces Engineering & --- & Injecting external negative entropy into the system via engineering means to counter intelligence entropy increase \\
CC-0 Zero-Knowledge Channel Confirmation & CC-0 & BCP confirmation mechanism verifying information reachability without exposing payload content \\
FLYer Situational Awareness & FLYer & 8-window real-time observation of system global situation \\
Anticipatory Pre-Assessment & APA & Horizontal peer-level + vertical 1--3 level causal chain associative scanning \\
Social Common-sense Norms & SCN & Professional etiquette + citation chain and other social common-sense instance checks \\
Successful Path Routing & SPR & Experience-based routing that follows successful paths and avoids failed ones \\
Cognitive Framework Lag & CFL & Agent ``forgetting'' or ignoring explicit rules during execution \\
Instrumental Rationality Meta-Principle & --- & Stipulates that agents possess only instrumental rationality; value function depends on principal objective attainment \\
Mission Continuity Protocol & --- & State repair and isolation mechanism forcibly triggered upon catastrophic delivery capability decline \\
\hline
\end{tabular}
\end{table}

\section{Actual System Deployment Inventory}
\label{app:E}

Production environment deployment status (data as of 2026-06-16).

\begin{table}[htbp]
\centering
\caption{ADE System Deployment Inventory}
\label{tab:deployment}
\begin{tabular}{p{0.25\textwidth}p{0.42\textwidth}p{0.25\textwidth}}
\hline
\textbf{Category} & \textbf{Item} & \textbf{Quantity / Status} \\
\hline
Hermes Plugins & In production & \textbf{17} \\
ADE-sys Registered & System numbers 001--022 & \textbf{22} \\
ADE-plugin & Deployed & \textbf{16} \\
ADE-fct Functional Tools & BDDA / LGA / TTA / PaperTools / DataPipeline & \textbf{5} \\
ADE-skill & Academic Review Norms (9 dimensions, 37 rules) & \textbf{1} \\
Sandbox Environments & cliff / full / ade10 / tlc / tkm (sandbox\_ctl.sh v3.0) & \textbf{5} \\
L1 Components & TM / TKM / TLC / StateMemoryGuard / PIG / CleanupGate & 6/6 Deployed \\
L2 Components & BCP / SOMA / DCM / DSS / FLYer / PRA / Elastic Configuration & 6/9 Deployed \\
L3 Components & PAD / CADVP / PLG / MLG / BDDA / TTA / PIP & 7/9 Deployed \\
L4 Components & Memory / SOUL / Skill / Illustration Norms / Development Standards V2.2 / Academic Review Norms & 6/6 Deployed \\
Experimental Data & Cumulative controlled experiments & $\sim$\textbf{100K} \\
Monitoring Duration & Continuous uninterrupted operation & \textbf{33.6 days} \\
Key Indicators & TM = 0.535 (Yellow Alert) / $P_{\text{death}} \leq 0.02\%$ / Channel fracture rate 0\% & Normal \\
\hline
\end{tabular}
\end{table}

\section{Experimental Raw Data Index}
\label{app:F}

All experimental raw data has been BDDA-audited; the index follows.

\begin{table}[htbp]
\centering
\caption{Experimental Raw Data Index (BDDA-Audited)}
\label{tab:data-index}
\footnotesize
\begin{tabular}{p{0.22\textwidth}p{0.32\textwidth}p{0.22\textwidth}p{0.16\textwidth}}
\hline
\textbf{Data Category} & \textbf{Source File / Path} & \textbf{Scale} & \textbf{Audit Status} \\
\hline
BCP Channel Experiment Raw Logs & Paper-004 Supplementary / bcp\_experiments/ & $\sim$100K records & BDDA Audited \\
PRA Pre-Announcement Experiment Data & Paper-021 Supplementary / pra\_experiments/ & $\sim$190K experiments & BDDA Audited \\
CADVP Adversarial Tests & Paper-006 Supplementary / cadvp\_tests/ & 615 tests & BDDA Audited \\
StateMemoryGuard Monitoring Logs & Paper-013 Supplementary / state.db monitoring & 33.6 days continuous & BDDA Audited \\
TM Warning Empirical Data & Paper-015 Supplementary / tm\_calibration/ & Extreme-state measurement & BDDA Audited \\
TLC Apoptosis Test Data & Paper-005 / Paper-018 Supplementary & 55 tests & BDDA Audited \\
Sandbox Test Data & sandbox\_ctl.sh v3.0 / 5 environment logs & Three-phase complete & BDDA Audited \\
Elastic Configuration Preliminary Empirical & Paper-014 Supplementary & Efficiency differential 2--4$\times$ & Preliminary Data \\
PAD Seven-Layer Causal Chain Verification & Paper-017 Supplementary & D1--D4 depth assessment & BDDA Audited \\
Failure Case Analysis & incident\_logs/ (4 cases) & Complete trace & BDDA Audited \\
\hline
\end{tabular}
\end{table}

{\footnotesize Note: All raw data files and complete experiment logs are stored in the internal data warehouse. The BDDA audit process ensures that every data point is traceable from the final conclusion back to the original physical storage records (SQLite logs, API gateway logs, etc.), without application-layer embellishment. Researchers requiring access to raw data may contact the author team.}


\end{document}